\documentclass[aps,prd,onecolumn,nofootinbib]{revtex4-2}
\usepackage{amsmath,amssymb,amsfonts}
\usepackage{booktabs,tabularx,array}
\usepackage{xcolor,graphicx}
\usepackage{braket}
\usepackage[normalem]{ulem}
\usepackage{hyperref}
\hypersetup{
colorlinks=true,
citecolor=blue,  
linkcolor=blue,  
urlcolor=blue    
}

\begin{document}

\title{Squeezing and Decoherence of Axion Dark Matter}

\author{Vasileios Fragkos$^{1,2}$}
\email{vfragkos@stevens.edu}
\author{Igor Pikovski$^{1,2}$}
\email{pikovski@stevens.edu}

\affiliation{$^{1}$ Department of Physics, Stevens Institute of Technology, Hoboken, New Jersey 07030, USA}
\affiliation{$^2$Department of Physics, Stockholm University, SE-106 91 Stockholm, Sweden}

\begin{abstract}
Bosonic particles beyond the Standard Model, such as axions, are among the leading candidates for dark matter. If such particles have a small mass, they are usually described as a purely classical field. However, it was recently shown that this description is incomplete: their inherent gravitational self-interactions can squeeze the field, producing a quantum feature beyond the classical description. Here we analyze squeezing of light dark matter in the presence of decoherence. We show that squeezing is continuously generated, and rapidly enough to survive dissipation, and may also remain present under thermal fluctuations and dephasing. Our results demonstrate that dark matter might possess persistent quantum features and motivate further investigation of their implications for dark matter detection.
\end{abstract}
\maketitle
\section{Introduction}

Axions and axion-like particles constitute one of the most motivated extensions of the Standard Model of particle physics and are among the leading candidates for dark matter \cite{PreskillWiseWilczek1983,AbbottSikivie1983,DineFischler1983}. Originally introduced in connection to the strong CP problem \cite{PecceiQuinn1977,Weinberg1978,Wilczek1978}, axions have since motivated a much broader phenomenological framework that includes ultra-light axion particles \cite{HuBarkanaGruzinov2000,HuiOstrikerTremaineWitten2017}, string-inspired models and the axiverse \cite{SvrcekWitten2006,ArvanitakiDimopoulosDubovskyKaloperMarchRussell2010} and coherent bosonic dark matter spanning many orders of magnitude in mass \cite{Turner1983,Marsh2016AxionCosmology,AriasCadamuroGoodsellJaeckelRedondoRingwald2012}.

Even though their couplings to ordinary matter and photons are extremely weak \cite{GrahamIrastorzaLamoreauxLindnerVanBibber2015}, they do provide the portals through which the axion and axion-like particles can be searched for \cite{Youdin1996,Aprile2014Axions,Abel2017Axions,Garcon2019CASPEr,Terrano2019Axions,Ouellet2019ABRACADABRA,Du2018ADMX,Anastassopoulos2017CAST,Carenza2020StimulatedDecay}. At the same time, their gravitational and cosmological effects provide complementary probes \cite{Hlozek2015Axions,Irsic2017FuzzyDM,Arvanitaki2015BlackHoles,Porayko2018PulsarTiming,Planck2018Inflation,Carenza2025AxionAstrophysics}. This has motivated large experimental and observational efforts, ranging from axion haloscopes \cite{Sikivie1983ExperimentalTests} and helioscopes \cite{VanBibber1989SolarAxions,CAST2007SolarAxions,IAXO2014ConceptualDesign} to precision interferometry \cite{GrahamRajendran2011ColdMolecules,GrahamRajendran2013NewObservables,ArvanitakiGrahamHoganRajendranVanTilburg2018}, astrophysical probes \cite{Raffelt1996StarsAsLaboratories,Raffelt2006AstrophysicalAxionBounds} and cosmological structure formation \cite{HuBarkanaGruzinov2000,HuiOstrikerTremaineWitten2017,Marsh2016AxionCosmology}.

For the purposes of this work, the most important aspect of axion dark matter is that it is usually treated as a highly occupied bosonic field \cite{Marsh2016AxionCosmology,Hui2021WaveDarkMatter,GuthHertzbergPrescodWeinstein2015}. For a representative QCD-axion mass, $m_a\sim 10^{-5} \rm eV/c^2$, the occupation number in a de Broglie volume can be enormous, of order $N_{\rm dB}\sim 10^{26}$, for typical galactic dark matter velocities \cite{Hui2021WaveDarkMatter}. In the ultralight mass regime, $m_a\sim 10^{-22} \rm eV/c^2$, the occupation number becomes even larger, potentially reaching $N_{\rm dB}\sim 10^{94}$ while the de Broglie wavelength reaches galactic scales $\lambda_{\rm dB}\equiv (h/m_a\upsilon)\simeq 0.1$--$1\,{\rm kpc}$ \cite{Hui2021WaveDarkMatter,SchiveLiaoWoo2014,Schive2014WaveDM,Eberhardt:2025caq}. This highly occupied wave-like regime motivates the common description of axion dark matter as a coherently oscillating \textit{classical field}, an assumption that underlies many recent proposals for ultralight dark matter detection \cite{Budker2014CASPEr,Sikivie2014LC,Dror2025Astrometric,Kalia2024LevitatedFerromagnets,ULDMwithTrappedatominterf,Manley2020MechanicalResonators}.

However, the classical description of the field may not be sufficient \cite{Sikivie:2009qn,SikivieTodarello2017,KoppFragkosPikovski2022,EberhardtZamoraKoppAbel2023MeanField,EberhardtZamoraKoppAbel2024ClassicalField}, even though it is deeply rooted in the standard treatment of ultra-light axion dark matter. 
 
In the usual argument, the axion field is highly occupied, $N\gg 1$, and therefore the size of the initial relative quantum fluctuations, under the coherent state assumption, is expected to scale as $N^{-1/2}$. The dynamics is then assumed to be captured by mean-field, classical equations of motion, such as the Gross-Pitaevskii or Schr\"{o}dinger-Newton equations \cite{GuthHertzbergPrescodWeinstein2015,WidrowKaiser1993,DvaliZell2018,Rindler-Daller:2011vpv}. This viewpoint provides the basis for much of the axion phenomenology. However, a large occupation number by itself does not provide a sufficient criterion for classicality. In quantum optics and quantum information science, classicality or non-classicality is a property of the quantum state and not only its mean occupation number, and quantum features can take many different forms. For instance, non-classicality is usually benchmarked through criteria such as the non-positivity of the Glauber-Sudarshan $P$-function, presence of quadrature squeezing, sub-Poissonian statistics, Wigner negativity and entanglement, among many others \cite{Glauber1963CoherentIncoherent,MandelWolf1995,Dodonov2002,BraunsteinVanLoock2005,Weedbrook2012,KenfackZyczkowski2004,Yadin2018,GoldbergKlimovGrassl2020}. For axions, large occupation only says that the initial fluctuations may be small. If the Hamiltonian that drives the dynamics is non-linear, as is the case for self-gravitating and self-interacting axions \cite{PreskillWiseWilczek1983,Erken2012}, these fluctuations can evolve non-trivially, even when the mean field itself continues to obey a classical equation of motion. This is familiar from quantum optics and cold-atom physics such as in Bose-Einstein condensates, which is often the analogy to ultra-light fields, where highly occupied fields evolving under non-linear dynamics give rise to squeezing and other non-classical signatures despite the fact that their mean field satisfies a classical equation of motion \cite{WallsMilburn2008,GreinerMandelHaenschBloch2002,OrzelTuchmanFenselauYasudaKasevich2001,ParkinsWalls1998}.

In previous work \cite{KoppFragkosPikovski2022}, we showed that  non-classical features in fact do arise naturally in ultra-light axion-like fields, where squeezing is generated on short time-scales: Keeping the quantum operator associated with the highly occupied axion mode, while imposing the Hartree/mean-field ansatz for the mode functions, the quantum operator obeys a non-linear Kerr-oscillator equation of motion \cite{YurkeStoler1986,YurkeStoler1988}. At the level of the mean field, the evolution appears classical. However, at the level of the fluctuations, the non-linear evolution squeezes the quantum state. As a result, an initial coherent state does not remain coherent under the evolution: the quadrature fluctuations rapidly fall below the coherent/vacuum noise level from just the  inherent gravitational dynamics of the field. In this sense, axion dark matter can develop a genuinely quantum feature, even in a regime usually regarded as classical. An analogue of this effect is well-known in Bose-Einstein condensates \cite{ParkinsWalls1998,OrzelTuchmanFenselauYasudaKasevich2001,Esteve2008}, which can exhibit macroscopic classical wave behavior and quantum features simultaneously.  For ultra-light axions, in addition to inherent squeezing from the non-linear evolution demonstrated in our work \cite{KoppFragkosPikovski2022}, other non-classical features have recently been proposed \cite{AllaliHertzberg2021,Lentz2025Nonclassical}.

Here, we investigate whether the rapid squeezing of axions can survive under decoherence.
The single-mode unitary analysis in \cite{KoppFragkosPikovski2022} establishes that gravitational self-interactions lead to squeezing. Here, we study its robustness against a range of different decoherence mechanisms that are likely to arise. This is especially important because decoherence is often invoked as the mechanism that restores classical behavior \cite{Zeh1970,Zurek1982,JoosZeh1985,Zurek2003,Schlosshauer2005}. If the environmental interactions destroy the squeezing on a timescale shorter than its onset, no observational consequences would be expected. On the other hand, if squeezing is re-generated before decoherence becomes dominant, or similarly if the quadrature noise remains below the vacuum plateau for a finite window of time, then quantum dynamics of axions may survive and may have observable consequences.  Here we show that indeed squeezing may remain robust against decoherence. We study a broad range of decoherence mechanisms, such as multi-mode interactions in the field, dissipation and dephasing into both the vacuum and a thermal bath. Our results show that the inherent non-linearity of the field evolution can be stronger than the decoherence, and that squeezing of axions may survive in the real world.

The paper is organized as follows. In section \ref{sec:2}, we summarize our previous findings on axion squeezing in \cite{KoppFragkosPikovski2022} for the unitary single-mode axion dynamics. We then proceed to analyze its robustness against decoherence through an open-system analysis. In section \ref{sec:Toymodel}, we introduce a two-mode axion toy model, illustrating the inter-mode couplings that generate entanglement, and show how tracing over one mode yields reduced-state dephasing. We highlight that the squeezing can survive and is robust in such a model. In section \ref{sec:Decoherence of ADM}, we study the effect of representative Markovian environments on Kerr squeezing, focusing on zero-temperature damping, dephasing and finite-temperature dissipation. We derive the corresponding expressions for the minimal quadrature variance and identify the regimes in which squeezing survives. Finally, we summarize our conclusions in section \ref{sec:conclusions} and present supporting derivations in the appendices. 

\section{Summary of Unitary Axion Dynamics}\label{sec:2}
In this section, we summarize our core results in \cite{KoppFragkosPikovski2022}. In a nutshell, Ref. \cite{KoppFragkosPikovski2022} showed that axion-like dark matter, although it is usually described as a classical field \cite{HuiOstrikerTremaineWitten2017,Marsh2016AxionCosmology,Hui2021WaveDarkMatter,GuthHertzbergPrescodWeinstein2015}, naturally develops a genuinely quantum mechanical feature through its gravitational interaction: quadrature squeezing below the coherent-state, or vacuum, level. The standard argument for treating axions as a classical field is based on their enormous occupation number \cite{GuthHertzbergPrescodWeinstein2015}. In the late Universe, many axions occupy the same macroscopic field configuration, and the relative size of quantum fluctuations is expected to scale as $N^{-1/2}$ \cite{GuthHertzbergPrescodWeinstein2015}. This motivates the usual Gross-Pitaevskii \cite{PitaevskiiStringari2003BEC} or Schr\"{o}dinger-Poisson description, where the axion field is effectively 
treated as a classical wave \cite{Hui2021WaveDarkMatter}. 

The main message of \cite{KoppFragkosPikovski2022} is that this classical-field intuition (or description) is incomplete. If one keeps the quantum operator associated with the macroscopically occupied mode, $\hat a$, the same gravitational self-interaction that appears in the Gross-Pitaevskii equation also generates, at the operator level, a Kerr-type quantum nonlinearity \cite{YurkeStoler1986}. 
This term,  intrinsic to the gravitational dynamics of the field, squeezes the quantum fluctuations around the mean field, as we briefly summarize below.

To briefly illustrate our finding, our starting point in \cite{KoppFragkosPikovski2022} is the Hartree approximation, in which the non-relativistic axion quantum field
is assumed to be dominated by a single mode,
\begin{equation}
\hat{\psi}(\mathbf{x},t)
\simeq
\psi(\mathbf{x},t)\frac{\hat a(t)}{\sqrt{N}} 
\end{equation}
Here $\psi(\mathbf{x},t)$ is normalized such that $\int_V d^3x|\psi(\boldsymbol{x},t)|^2=N$ and obeys the usual mean-field Schr\"{o}dinger-Poisson dynamics, while $\hat a(t)$ remains a genuine annihilation operator satisfying $[\hat a,\hat a^\dagger]=1 $. The naive classical approximation corresponds to replacing 
\begin{equation}
\hat a(t)\rightarrow \sqrt{N}
\end{equation}
This replacement gives the standard classical field description, but \textit{discards the quantum fluctuations/noise of the axion field}. The key observation of our work \cite{KoppFragkosPikovski2022} is that this discarded quantum noise is \textit{not passive but is dynamically squeezed by the self-gravity of the axion field}.

\begin{figure}[t]
\centering
\includegraphics[
width=0.80\linewidth,
height=0.3\textheight,keepaspectratio
]{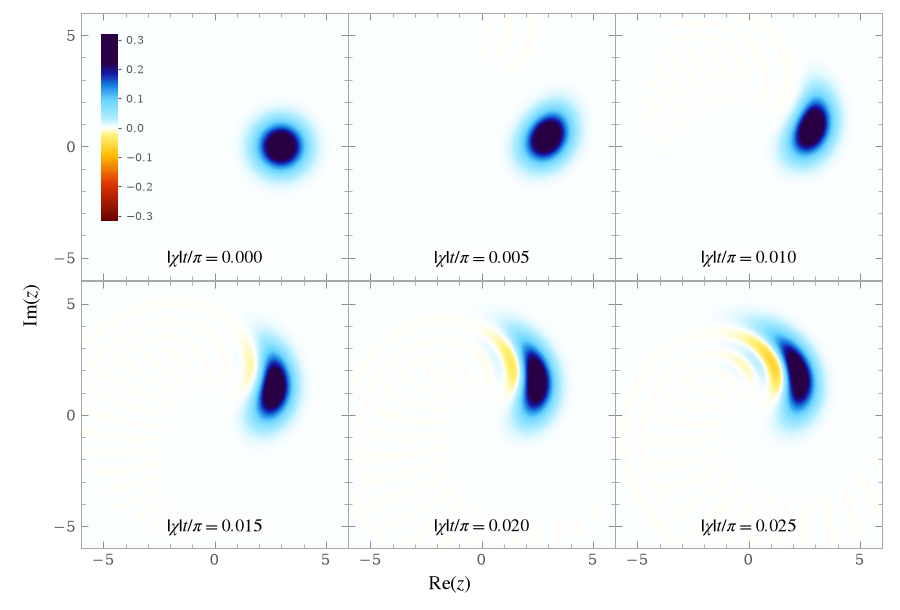}
\caption{ Evolution of the Wigner function $W(z,t)=\frac{2}{\pi}Tr\bigl[\hat{\rho}(t)\hat{D}(z)(-1)^{\hat{n}}\hat{D}^{\dagger}(z)\bigr]$ with $\hat{D}(z)\equiv e^{z\hat{a}^{\dagger}-z^*\hat{a}}$ of an initially coherent state $\hat{D}(3)\ket{0}$ under the Kerr Hamiltonian \eqref{kerrHamiltonian}.  Here, $|\chi|$ is the strength of Kerr non-linearity.
The figure shows the early-time nonlinear shearing and quadrature squeezing. At later times, the full unitary Kerr evolution includes interference fringes, cat-like states, and revival \cite{YurkeStoler1986,Buzek1989Periodic}. Here, we are focusing only on squeezing, before the onset of other coherent effects, which however is already a signature of quantum behavior.}
\label{fig:squeezingKerr}
\end{figure}

The dynamics of the operator $\hat{a}(t)$ takes the form of a well studied model in quantum optics as well as in cold atoms, the so-called Kerr oscillator \cite{ParkinsWalls1998,YurkeStoler1986,Tombesi1987Generation,YurkeStoler1988,agarwal1989quantum,TanasKielich1983,Buzek1989Periodic,miranowicz1990generation,OrzelTuchmanFenselauYasudaKasevich2001,GreinerMandelHaenschBloch2002} 
\begin{equation}\label{eq:Kerr:eom}
i\hbar\frac{d\hat a}{dt}=\hbar\omega(t)\hat a+2\hbar\chi(t)\hat a^\dagger \hat a\hat a 
\end{equation}
which is generated by a non-linear Hamiltonian of the form\footnote{As we argue in \cite{KoppFragkosPikovski2022}, $\omega(t),\chi(t)$ can be well approximated as frozen in time. During the formation of squeezing, the fractional changes $\frac{\dot \omega}{\omega}\approx 0$ as well as $\frac{\dot \chi}{\chi}\approx 0$.}
\begin{equation}\label{kerrHamiltonian}
\hat H= \hbar \omega \hat a^\dagger \hat a+
\hbar \chi \hat a^{\dagger 2}\hat a^2 
\end{equation}
where $\chi$ is the strength of the non-linearity\footnote{The naive substitution $\hat a(t)\rightarrow \sqrt{N}$ \textit{trivially} satisfies Eq. (\ref{eq:Kerr:eom}) within the self-consistent Hartree ansatz which, crucially, implies $\omega+2N\chi=0$. However, as stated in the main text, it neglects the quantum noise that undergoes strong squeezing. For a detailed analysis the reader is referred to Eqs. (3a)-(3f) and Appendix E of Ref. \cite{KoppFragkosPikovski2022}.}. The non-linear Kerr term makes different number components of the quantum state rotate slightly differently. As a result of this, in phase space, an initial circular coherent-state uncertainty distribution is `twisted', as illustrated in figure \ref{fig:squeezingKerr}. The effect to lowest order is that the variance of one quadrature, which we call minimal variance or principal squeezing \cite{TanasKielich1983,Walls1983SqueezedStates,TanasMiranowiczKielich1991,BajerMiranowiczTanas2002}, decreases the vacuum level, while its conjugate variance increases (at later times, the field becomes even more non-classical and the Kerr-nonlinearity generates large Schr\"{o}dinger-cat-states, but here we focus on the most robust quantum effect of squeezing). Defining $\hat X_\theta\equiv\hat a e^{-i\theta}+\hat a^\dagger e^{i\theta}$ \cite{WallsMilburn2008}, squeezing occurs when the minimal quadrature variance, in the Heisenberg picture, satisfies
\begin{equation}\label{eq:minimalvarianceDefinition}
V_-(t)=\min_\theta \,{\rm Var}\!\left(\hat X_\theta(t)\right)<1 \, .
\end{equation}
\\
As an initial state, the axion field is taken to begin in a coherent state $\ket{\alpha_0=\sqrt{N}}$. While this might be expected for a classical field, this is not an ad-hoc choice, instead it is physically motivated by the so-called misalignment mechanism \cite{PreskillWiseWilczek1983,AbbottSikivie1983,DineFischler1983}. In the early Universe, the axion field is assumed to be initially displaced from the minimum of its potential. While the Hubble rate is larger than the axion mass $m_a$, the field is approximately frozen by Hubble friction. Later, when the Hubble rate becomes smaller than the mass, $H(t)\lesssim m_ac^2/\hbar$, the field begins to oscillate coherently around the minimum of its potential \cite{Marsh2016AxionCosmology}.
Classically, this gives a homogeneous oscillating scalar field. Quantum
mechanically, the state that most closely represents a classical field with a
well-defined amplitude and phase is a coherent state \cite{Glauber1963CoherentIncoherent}. Therefore the coherent
state is not an arbitrary choice: it is the natural quantum representative of
the classical field produced by misalignment. For a coherent state, the relative field fluctuations $\delta \psi$ of the occupied mode scale as \cite{GuthHertzbergPrescodWeinstein2015},
\begin{equation}
\frac{\delta \psi}{\psi}\sim \frac{1}{\sqrt{N}} \, .
\end{equation}
For a large occupation number, $N\gg 1$, that is typical for axions \cite{Hui2021WaveDarkMatter}, we get $\frac{\delta \psi}{\psi}\rightarrow 0$. This is the standard classicality argument of axion dark matter. However, as we show in \cite{KoppFragkosPikovski2022}, even if the field initially looks classical, the Kerr evolution \eqref{eq:Kerr:eom} subsequently turns unitarily this apparently classical state into a squeezed state. 

Although an initial coherent state is physically motivated in the context of axion dark matter, we should stress it is not essential for the squeezing mechanism described in Ref.\cite{KoppFragkosPikovski2022}. The crucial aspect in our analysis is that the Kerr evolution operator, around a mean field, can be approximated as an effective squeezing operator (see Appendix $C$ in Ref.\cite{KoppFragkosPikovski2022}). As a result, reduced quadrature fluctuations for a broader class of initial states are possible, including displaced thermal states. However, in that case, going below the vacuum level would need stronger squeezing due to the additional initial noise of the state. In what follows, we focus on an initially coherent state. 

\begin{figure}[t]
\centering
\includegraphics[
width=0.9\linewidth,
height=0.3\textheight,keepaspectratio
]{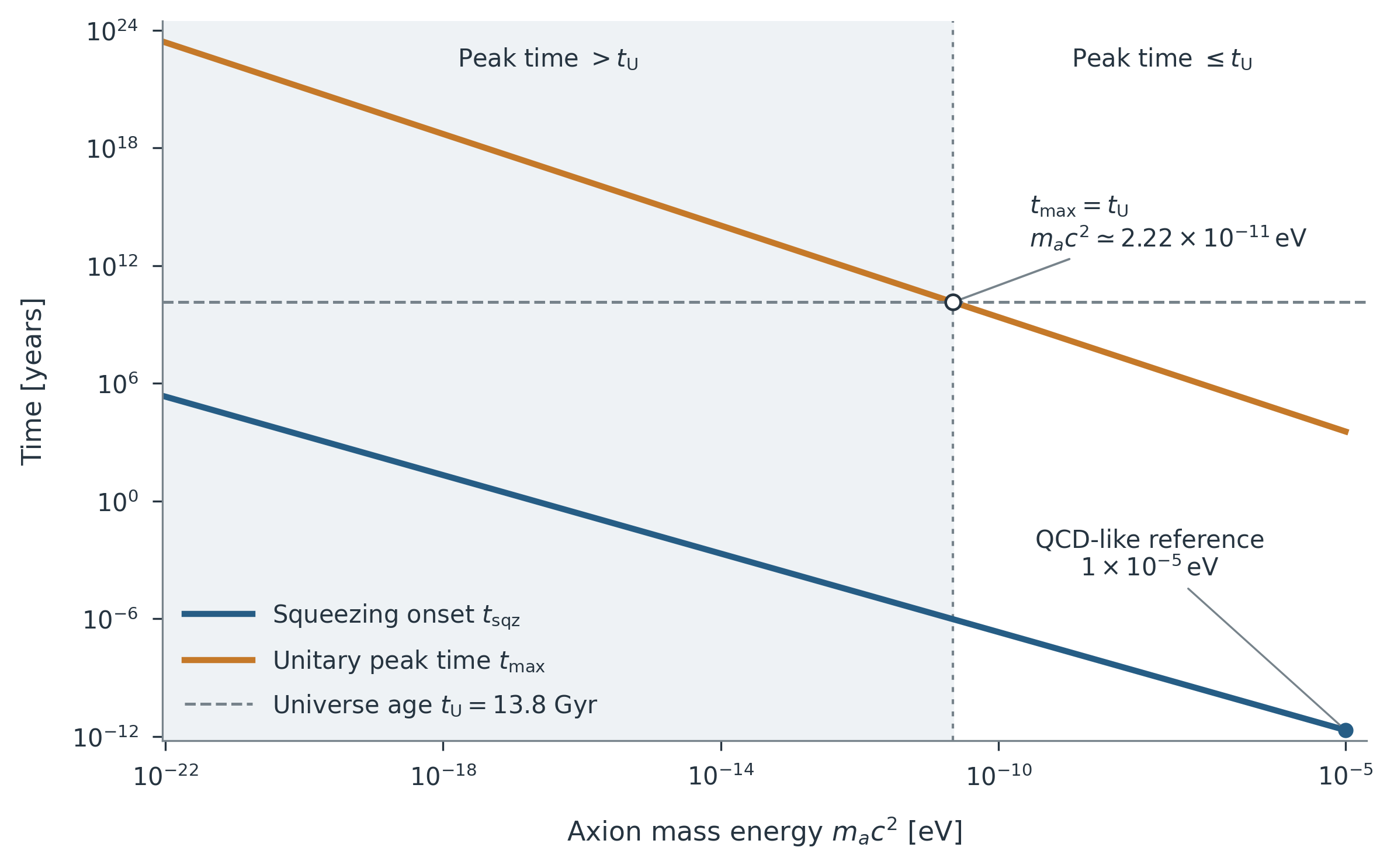}
\caption{\small{
Cosmological squeezing timescales as functions of the axion mass-energy.
The blue curve shows the onset time $t_{\rm sqz}\propto m_a^{-1}$, while the orange
curve shows the time $t_{\max}\propto m_a^{-7/6}$ at which maximal squeezing is reached.
The dashed horizontal line denotes the age of the Universe. It is clear that over the plotted axion mass range, the onset of squeezing is well below the age of the Universe.}}
\label{fig:cosmological_squeezing_timescales}
\end{figure}

Given the inherent non-linear axion dynamics, squeezing occurs for arbitrarily small positive time $t$. In \cite{KoppFragkosPikovski2022}, we operationally define the squeezing timescale $t_{\mathrm{sqz}}$ as the time required to reach $r(t_{\mathrm{sqz}})\equiv -\frac{1}{2}\ln V_-(t_{\mathrm{sqz}})=1$. In this model,this gives
\begin{equation}\label{unitarytsqz}
t_{\rm sqz}\simeq\frac{0.6}{N|\chi|} \, .
\end{equation}
At first sight, this appears to depend on the number of axions $N$. However, for the self-gravitating axion mode considered in \cite{KoppFragkosPikovski2022}, the Kerr coupling parameter scales as $|\chi| \propto \frac{1}{N}$.
\begin{figure}[t]
\centering
\includegraphics[
width=1.1\linewidth,
height=0.33\textheight,keepaspectratio
]{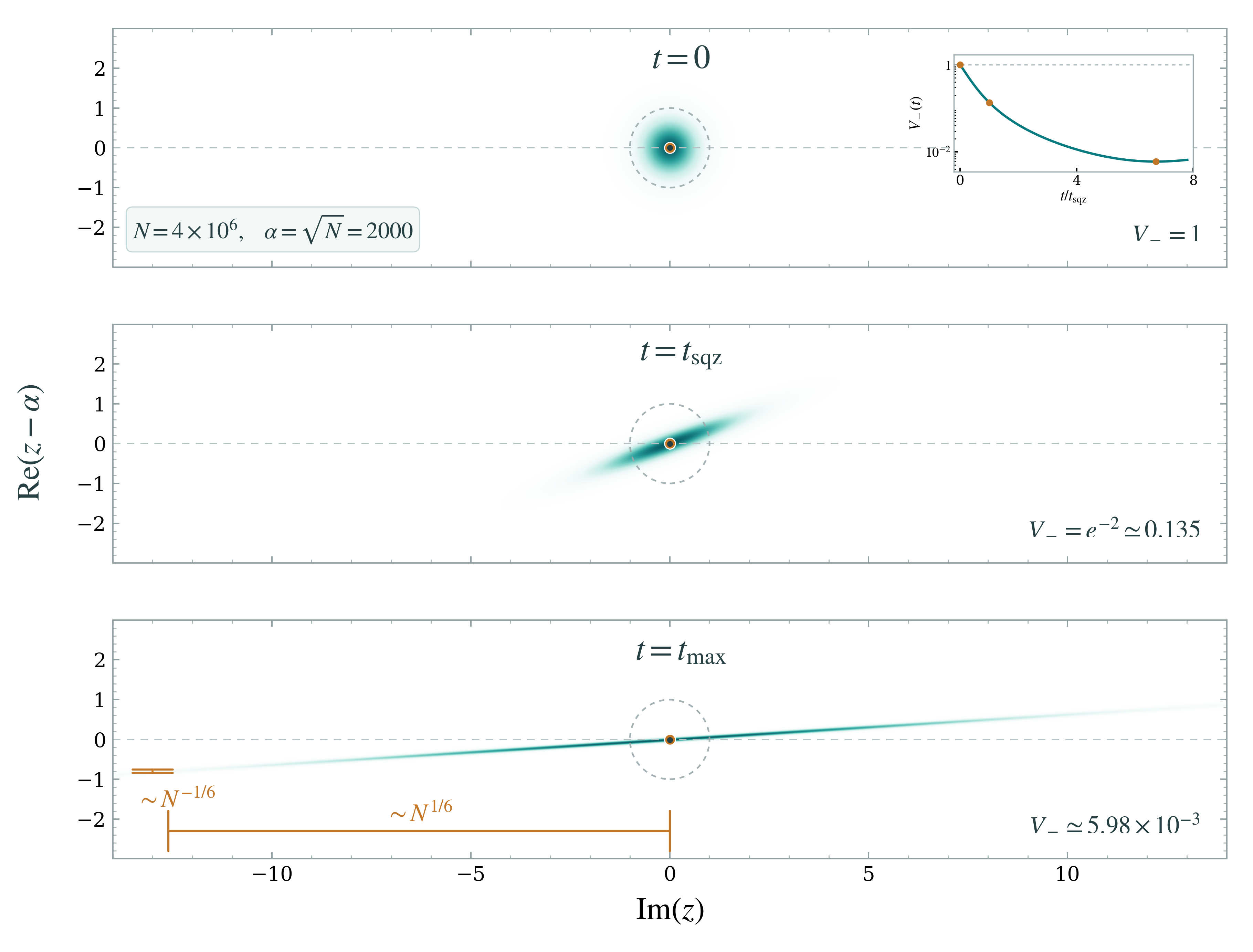}
\caption{Squeezing of axion dark matter as found in \cite{KoppFragkosPikovski2022}. An initial coherent state develops quadrature squeezing that results in the reduction of its quantum noise below the vacuum level  ${\rm Var}\!\left(\hat X_\theta(t)\right)=e^{-2r(t)}$, reaching $r=1$, at $t_{\rm sqz}\simeq \omega^{-1}\simeq (N|\chi|)^{-1}$. The conjugate quadrature is anti-squeezed with a corresponding exponential increase in its noise. Maximal squeezing is reached at $t_{\max}\simeq0.3N^{1/6}/(N|\chi|)$. The minimal quadrature variance starts at $V_-(0)=1$ and reaches $V_-(t_{\rm sqz})=e^{-2}$ at $t_{\rm sqz}$. Finally, at $t=t_{\rm max}$, the variance is $V_-(t_{\max})\simeq0.95N^{-1/3}$ \cite{KoppFragkosPikovski2022}. The inset shows the exact minimal quadrature variance $V_-(t)$, normalized to the vacuum value of one, with the three stages marked. }
\label{fig:squeezingcosmology}
\end{figure}
This is one of the most important conclusions in our previous work: The onset of squeezing is \textit{independent of the number of axions $N$} and does \textit{not} become slower for larger occupation number.\footnote{This finding evades standard textbook arguments that claim for $N\gg 1$, a quantum field can be described classically \cite{Sakurai1967AdvancedQM}. Large occupation suppresses initial relative fluctuations, but it does not by itself guarantee that the quantum state remains coherent under nonlinear evolution. Of course one could insist on a classical description, that captures some effects, but at the cost of concealing other interesting quantum effects.} Large $N$ makes the mean field look classical, but it \textit{does not delay the quantum squeezing} of the fluctuations around that mean field. For cosmological axion dark matter, we have estimated the squeezing time (see eq. 12 in \cite{KoppFragkosPikovski2022})
\begin{equation}
t_{\rm sqz}\sim\frac{\hbar}{m_a v_{\rm rms}^2}\sim\frac{\lambda_{\rm deB}}{v_{\rm rms}}\simeq \frac{10^{-5}\mathrm{eV}}{m_ac^2}66\mu\mathrm{ s}
\end{equation}
where $v_{\rm rms}\sim10^{-3}c$ is the typical axion velocity for virialized cosmological axions in galactic halos \cite{Hui2021WaveDarkMatter} and $\lambda_{\rm deB}$ is the reduced
de Broglie wavelength. For a QCD axion with a theoretically expected mass $m_ac^2\sim 10^{-5}\,{\rm eV}$ the result is of the order of
\begin{equation}
t_{\rm sqz}\simeq 66 \mu\rm s
\end{equation}
Thus, squeezing is generated on extremely short timescales compared with cosmological evolution. In particular, as illustrated in Fig. \eqref{fig:cosmological_squeezing_timescales}, for axions in the mass-energy range $10^{-22}\mathrm{eV}\leq m_ac^2\leq 10^{-5} \mathrm{eV}$, including ultralight axion dark matter candidates \cite{HuiOstrikerTremaineWitten2017}, the squeezing timescale turns out to be cosmologically relevant.

Unitary squeezing can grow until it reaches its first maximum at $t=t_{\rm max}$
\begin{equation}
r_{\max} \simeq \ln\!\big(N^{1/6}\big) 
\end{equation}
Here $N$ denotes the occupation of the single Hartree mode over the selected volume $V$. For the cosmological benchmark of Ref. \cite{KoppFragkosPikovski2022}, $V$ was chosen to be of order the present Hubble volume. For $N\sim 10^{93}$ axions in a Hubble volume, this yields $r_{\max}\approx 36$, corresponding to a remarkable $\sim313$~dB of squeezing.\footnote{For a squeezed variance, $S_{\rm dB}\equiv-10\log_{10}(e^{-2r})\simeq 8.686\cdot r$. For comparison, the largest directly measured optical quadrature squeezing in the laboratory is approximately $\sim 15\mathrm{dB}$ \cite{VahlbruchMehmetDanzmannSchnabel2016}. Inflationary perturbations \cite{Baumann2009Inflation} are often described as highly squeezed two-mode quantum states \cite{GrishchukSidorov1989,GrishchukSidorov1990,AlbrechtFerreiraJoyceProkopec1994}, with typical values, by the end of inflation, $r\sim 50$ or approximately $S_{\rm dB}\simeq434 \rm$ dB. This comparison is meant only as an indication of the enormous magnitude of the squeezing that axions can reach.} The time required to reach this maximum is  
\begin{equation}\label{maximumsqztime}
t_{\max} \approx 3500 \cdot \biggl(\frac{10^{-5}\mathrm{eV}}{m_ac^2}\biggr)^{\frac{7}{6}}\mathrm{yr}
\end{equation}
For QCD-like axions with $m_ac^2\approx10^{-5} \mathrm{eV}$, the value of $t_{\mathrm{max}}=3500\mathrm{yr}$ is still much shorter than the age of the Universe. On the other hand, for sufficiently ultralight axions $m_a\leq 2.2 \cdot10^{-11}\mathrm{eV}/c^2$, the onset of squeezing $t_{\mathrm{sqz}}$ can still be fast, but the system would not reach the maximal squeezing within the age of the Universe. 

Thus, the conclusion is that the intrinsic self-interactions contained in the axion field dynamics generate squeezing on extremely short time-scales. It can reach enormous values, but the overall size of the effect depends on the axion mass. Therefore, one might expect that ultra-light axion dark matter evolves rapidly into a quantum state, and that it could be highly squeezed in the process.
In addition to just the dynamics of the field in a cosmological setting, we have also considered other self-gravitating situations, such as
galactic halos and solitonic cores \cite{SchiveLiaoWoo2014}, and found the same qualitative result: the onset of squeezing is extremely rapid compared with astrophysical timescales.
\begin{figure}[t]
\centering
\includegraphics[
width=0.85\linewidth,
height=0.3\textheight,keepaspectratio
]{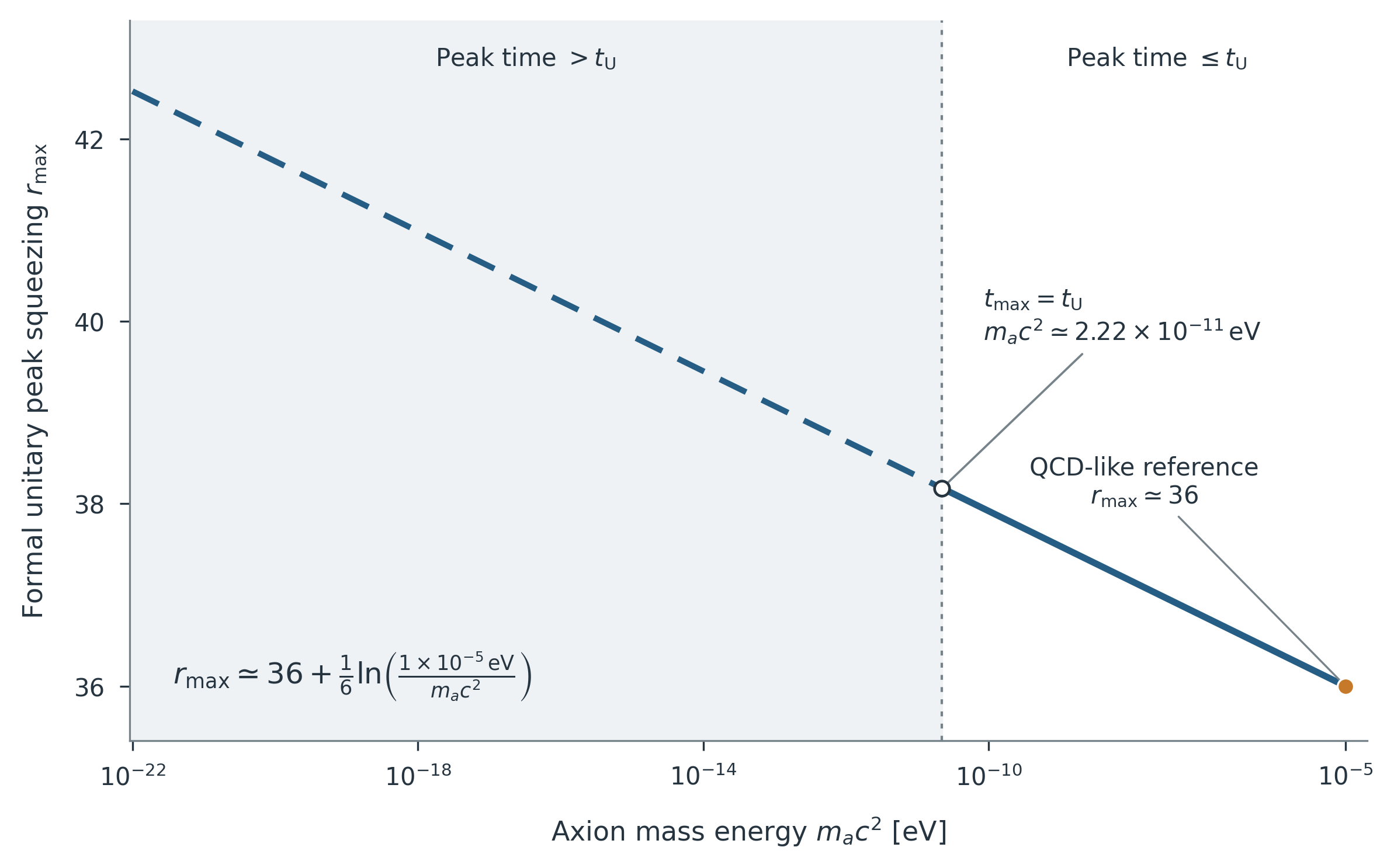}
\caption{\small
Maximum attainable squeezing $r_{\max}$ of the cosmological axion mode as a function of the axion mass-energy. The curve shows the estimate
$r_{\max}\simeq 36+\frac{1}{6}\ln(\frac{10^{-5}\mathrm{eV}}{m_ac^2})$ in \cite{KoppFragkosPikovski2022}, obtained from the large occupation number of axions in a Hubble volume, $N\simeq 10^{93} \cdot\frac{10^{-5}\mathrm{eV}}{m_ac^2}$. For the mass of the QCD axion, $m_a=10^{-5}\mathrm{eV}/c^2$, the value of maximum squeezing is $r_{\max}\simeq 36$. The vertical dotted line indicates the mass for which the time to reach maximal squeezing equals the age of the Universe, $t_{\max}=t_U$. For larger masses, $m_a\geq 2.2\cdot 10^{-11}\mathrm{eV}/c^2$ the maximal squeezing is reached within the age of the Universe.}
\label{fig:tmax}
\end{figure}
This result challenges the standard view that the axion field can simply be regarded as a coherent classical field. At the same time, our analysis is not outside the standard axion framework. It is precisely the Hartree approximation that leads to the Kerr oscillator for the occupied mode. In this sense, the squeezing is a quantum effect sitting on top of the usual mean-field description, and its description arises consistently within the same framework. We also note that this exact behavior is also expected in Bose-Einstein condensates \cite{ParkinsWalls1998,SalazarBECsqueezing}.

However, these results are only valid for a closed-system description in which the field evolves unitarily. Most importantly, the derivation assumes unitary evolution of a simple single-mode axion system, neglecting both environmental decoherence and possible multi-mode effects of the field. It is thus natural to ask whether squeezing may be maintained in a real-world scenario, or whether decoherence completely destroys its appearance.

In the following, we go beyond this idealized unitary picture, retaining the Kerr squeezing mechanism, while embedding the single axion mode in an open-system setting. In what follows, we will study how the minimal variance evolves under zero-temperature dissipation, finite-temperature dissipation, and pure dephasing. The central question then becomes a competition between the gravitational squeezing scale, controlled by $N|\chi|$, and the relevant environmental rates. Ultimately, this open-system extension will inform us on how much of the gravitationally generated axion squeezing survives once the ideal single-mode picture is embedded in a more realistic late-Universe environment. As we show below, our results suggest that the expected squeezing may be robust even in the presence of interactions with an environment.

\section{Two-mode axion dynamics, entanglement and reduced state dephasing}\label{sec:Toymodel}

In \cite{KoppFragkosPikovski2022}, we considered the axion dark matter field consisting of a single mode. The dynamics were treated within the Hartree approximation, which guarantees the validity of the Gross-Pitaevskii equation while retaining the full quantum operator nature of the field operators. That is, the non-relativistic axion field operator is $\hat{\psi}(\boldsymbol{x},t)=\frac{1}{\sqrt{N}}\psi(\boldsymbol{x},t)\hat{a}(t)$, keeping $\hat{a}(t)$ intact. This approach led to a system of equations that combines the usual Gross-Pitaevskii dynamics for $\psi(\boldsymbol{x},t)$ with a non-linear Kerr oscillator equation of motion for $\hat{a}(t)$, which, when acting on an initially coherent state, produces large quadrature squeezing.

Here, we relax the single-mode restriction and consider the case where axion dark matter comprises \textit{two interacting modes}. This provides an analytically tractable model for examining how the neglected degrees of freedom in the single-mode description affect squeezing. Our motivation is to show that different inter‑mode couplings give rise to richer axion dynamics, such as entanglement between modes and that the standard assumption of classicality may break down even for very simple two‑mode models. 

In subsection \ref{Multimode Hamiltonian}, we present the multimode Hamiltonian for non-relativistic, gravitationally interacting axions. In \ref{Subsection:2-mode case} we develop the two-mode truncation and analyze the exactly solvable dynamics which include cross and self-Kerr interactions. We compute the reduced density matrix of the selected mode $a$, the minimal quadrature variance and the linear entropy. Finally, in \ref{sec:bridgingtwomodeLindblad}, we briefly introduce the Markovian Lindblad framework for more general environments, which will prepare the broader analysis in Sec.\ref{sec:Decoherence of ADM}. 

\subsection{Multimode Hamiltonian for Non-Relativistic Gravitationally Interacting Axions}\label{Multimode Hamiltonian}

We consider a non-relativistic axion field interacting through Newtonian gravity, neglecting non-gravitational self-interactions and external potentials. From this section onward, including Sec. \ref{sec:Decoherence of ADM}, we use the frequency Hamiltonian $\hat H\equiv\hat H_{\mathrm{phys}}/\hbar$. The corresponding Hamiltonian reads \cite{GuthHertzbergPrescodWeinstein2015,Chakrabarty2018Gravitational,ArzaChakrabartyEnomotoHanTodarello2019}
\begin{equation}
\hat{H}=\frac{\hbar}{2m}\int d^3\boldsymbol{x}\nabla\hat{\psi}^{\dagger}(\boldsymbol{x},t)\cdot \nabla \hat{\psi} (\boldsymbol{x},t)  -\frac{Gm^2}{2\hbar}\int d^3\boldsymbol{x} \int d^3\boldsymbol{x}^{\prime} \frac{\hat{\psi}^{\dagger}(\boldsymbol{x},t)\hat{\psi}^{\dagger}(\boldsymbol{x}^{\prime},t)\hat{\psi}(\boldsymbol{x},t)\hat{\psi}(\boldsymbol{x}^{\prime},t)}{|\boldsymbol{x}-\boldsymbol{x}^{\prime}|}
\end{equation}
The Heisenberg equation of motion for the non-relativistic axion field $\hat{\psi}$ is\footnote{In units $\hbar=c=1$, the non-relativistic axion field $\hat{\psi}(\boldsymbol{x},t)$ is obtained by $\hat{\phi}(\boldsymbol{x},t)=\bigl(\hat{\psi}(\boldsymbol{x},t)e^{-imt}+\hat{\psi}^{\dagger}(\boldsymbol{x},t)e^{imt}\bigr)/\sqrt{2m}$, where $\hat{\phi}(\boldsymbol{x},t)$ is the fully relativistic axion scalar field. For a more detailed analysis the reader is referred to Ref. \cite{Namjoo:2017nia}.},\footnote{From now on, the axion mass is denoted by $m$.}
\begin{align*}
i\partial_t\hat{\psi}(\boldsymbol{x},t)=[\hat{\psi}(\boldsymbol{x},t),\hat{H}]=-\frac{\hbar}{2m}\nabla^2\hat{\psi}(\boldsymbol{x},t)+\frac{m}{\hbar}\hat{\Phi}(\boldsymbol{x},t)\hat{\psi}(\boldsymbol{x},t)
\end{align*}
where 
\begin{equation}
\hat{\Phi}(\boldsymbol{x},t)\equiv-Gm\int d^3\boldsymbol{x^{\prime}}\frac{\hat{\psi}^{\dagger}(\boldsymbol{x^{\prime}},t)\hat{\psi}(\boldsymbol{x^{\prime}},t)}{|\boldsymbol{x}-\boldsymbol{x^{\prime}}|}
\end{equation}
is the Newtonian potential, satisfying the Poisson equation $\nabla^2\hat{\Phi}(\boldsymbol{x},t)=4\pi Gm\hat{\psi}^{\dagger}(\boldsymbol{x},t)\hat{\psi}(\boldsymbol{x},t)$. We expand the axion field with respect to a complete orthonormal set of mode functions $u^{\vec{\alpha}}(\boldsymbol{x},t)$, that is $\hat{\psi}(\boldsymbol{x},t)=\sum_{\vec{\alpha}}u^{\vec{\alpha}}(\boldsymbol{x},t)\hat{a}_{\vec{\alpha}}(t)$ where $ \int_V d^3 \boldsymbol{x} u^{*\vec{\alpha}}(\boldsymbol{x},t)u^{\vec{\beta}}(\boldsymbol{x},t)=\delta^{\vec{\alpha}}_{\vec{\beta}}$ and $\sum_{\vec{\alpha}}u^{\vec{\alpha}}(\boldsymbol{x},t)u^{*\vec{\alpha}}(\boldsymbol{x^{\prime}},t)=\delta^{3}(\boldsymbol{x}-\boldsymbol{x^{\prime}})$ \cite{Chakrabarty2018Gravitational}. The equation of motion for the annihilation operator $\hat{a}_{\vec{\alpha}}(t)$ is 
\begin{equation}\label{eq:HeisenbergEOM}
i\partial_t \hat{a}_{\vec{\alpha}}=\sum_{\vec{\beta}}M_{\vec{\alpha}}^{\vec{\beta}}\hat{a}_{\vec{\beta}}+\frac{1}{2}\sum_{\vec{\beta},\vec{\gamma},\vec{\delta}}\Lambda_{\vec{\alpha},\vec{\beta}}^{\vec{\gamma},\vec{\delta}}\hat{a}_{\vec{\beta}}^{\dagger}\hat{a}_{\vec{\gamma}}\hat{a}_{\vec{\delta}}
\end{equation}
where the kinetic matrix elements are 
\begin{equation}\label{KineticMatrix}
M_{\vec{\alpha}}^{\vec{\beta}}=\int_V d^3x u^{*\vec{\alpha}}(\boldsymbol{x},t)\bigl(-i\partial_t-\frac{\hbar}{2m}\nabla^2\bigr)u^{\vec{\beta}}(\boldsymbol{x},t)     
\end{equation}
and the gravitational self-couplings 
\begin{equation}\label{LambdaCouplings}
\Lambda_{\vec{\alpha},\vec{\beta}}^{\vec{\gamma},\vec{\delta}}=-\frac{Gm^2}{\hbar}\int_V\int_V d^3\boldsymbol{x} d^3\boldsymbol{x^{\prime}} \frac{u^{*\vec{\alpha}}(\boldsymbol{x},t)   u^{*\vec{\beta}}(\boldsymbol{x^{\prime}},t)   \bigl(  u^{\vec{\gamma}}(\boldsymbol{x},t)    u^{\vec{\delta}}(\boldsymbol{x^{\prime}},t)+ u^{\vec{\gamma}}(\boldsymbol{x^{\prime}},t)    u^{\vec{\delta}}(\boldsymbol{x},t)\bigr)  }{|\boldsymbol{x}-\boldsymbol{x^{\prime}}|}    
\end{equation}
The Hamiltonian, in terms of creation and annihilation operators, in full generality, is \cite{Chakrabarty2018Gravitational,lentz2017improving}
\begin{equation}\label{GeneralHami}
\hat{H} = \sum_{\vec{\mu},\vec{\nu}} M_{\vec{\mu}}^{\vec{\nu}} \hat{a}_{\vec{\mu}}^{\dagger} \hat{a}_{\vec{\nu}} + \frac{1}{4} \sum_{\vec{\mu},\vec{\nu},\vec{\rho},\vec{\sigma}} \Lambda_{\vec{\mu},\vec{\nu}}^{\vec{\rho},\vec{\sigma}} \hat{a}_{\vec{\mu}}^{\dagger} \hat{a}_{\vec{\nu}}^{\dagger} \hat{a}_{\vec{\rho}} \hat{a}_{\vec{\sigma}}
\end{equation}
The matrix elements satisfy $M_{\vec{\alpha}}^{\vec{\beta}}=\bigl(M^{\vec{\alpha}}_{\vec{\beta}}\bigr)^*$ and $\Lambda_{\vec{\alpha},\vec{\beta}}^{\vec{\gamma},\vec{\delta}}=\Lambda_{\vec{\beta},\vec{\alpha}}^{\vec{\gamma},\vec{\delta}}=\Lambda_{\vec{\alpha},\vec{\beta}}^{\vec{\delta},\vec{\gamma}}=\bigl(\Lambda^{\vec{\alpha},\vec{\beta}}_{\vec{\gamma},\vec{\delta}}\bigr)^*$. This mode representation makes explicit the dynamics and interactions of a selected axion mode. The coefficients $\Lambda_{\mu\nu}^{\rho\sigma}$ appearing in the multi-mode Hamiltonian \eqref{GeneralHami} arise from the inherent gravitational interactions and include self-Kerr, Cross-Kerr and other processes that redistribute axion particles between modes. Their magnitudes, in the case of a single self-gravitating axion dark matter field, without any other external environment, are fixed by the spatial profiles and normalization of the modes. Thus, Hamiltonian \eqref{GeneralHami} lists the interaction processes that are, in principle, allowed by its operator structure but does not imply that every term is non-zero for every choice of modes.

A single orbital/mode function does not, by itself, determine the quantum state of the associated mode: Its quantum fluctuations can evolve non-trivially, as in the case of Kerr dynamics discussed in Sec.\ref{sec:2} and in \cite{KoppFragkosPikovski2022}. Allowing additional occupied modes, introduces the possibility of inter-mode interactions. Systematic treatments of such dynamics, are well studied in the context of Bose-Einstein Condensates, including multi-configurational\footnote{In this framework, the usual Gross-Pitaevskii description corresponds to the special Hartree limit in which all axions occupy a single mode. Beyond this limit, the dynamics become complicated since the evolution of the modes is governed by a system of $M$ coupled Gross--Pitaevskii-like equations \cite{Sakmann2011}.} approaches \cite{Sakmann2011}.
\subsection{Two-mode axion dynamics}\label{Subsection:2-mode case}
We now restrict the axion field expansion to two orthonormal modes. This truncation provides a more tractable setting in which we can identify how the inter-mode interactions affect the quantum state of a selected mode, and therefore its squeezing. Our aim in this subsection is to illustrate the interplay between the self-Kerr squeezing and the coupling to an unobserved mode, without attempting a complete description of the multi-mode axion dynamics. In this setting, the non-relativistic axion field operator takes the form 
\begin{equation}
\hat{\psi}(\boldsymbol{x},t)\simeq u^{(a)}(\boldsymbol{x},t)\hat{a}(t)+u^{(b)}(\boldsymbol{x},t)\hat{b}(t)
\end{equation}
where $\hat{a}(t),\hat{b}(t)$ annihilate an axion in the modes $u^{(a)}(\boldsymbol{x},t)$ and $u^{(b)}(\boldsymbol{x},t)$ respectively. The Hamiltonian in such a case is 
\begin{equation}\label{eq:twomodeHami}
\begin{aligned}
\hat{H} &= \underbrace{ M_{0}^{0}\,\hat{a}^{\dagger}\hat{a} + M_{1}^{1}\,\hat{b}^{\dagger}\hat{b}}_{\text{Free evolution}} + \underbrace{M_{0}^{1}\,\hat{a}^{\dagger}\hat{b} 
+ M_{1}^{0}\,\hat{b}^{\dagger}\hat{a} }_{\text{Single axion hopping}} + \underbrace{\frac{1}{4}\Lambda_{00}^{00}\,\hat{a}^{\dagger}\hat{a}^{\dagger}\hat{a}\hat{a}+ \frac{1}{4}\Lambda_{11}^{11}\,\hat{b}^{\dagger}\hat{b}^{\dagger}\hat{b}\hat{b}}_{\text{Self-Kerr}}+ \underbrace{\Lambda_{01}^{01}\,\hat{a}^{\dagger}\hat{b}^{\dagger}\hat{a}\hat{b}}_{\text{Cross-Kerr}} \\
&\quad + \underbrace{\frac{1}{2}\Lambda_{01}^{00}\,\hat{a}^{\dagger}\hat{b}^{\dagger}\hat{a}\hat{a}
+\frac{1}{2}\Lambda_{01}^{11}\,\hat{a}^{\dagger}\hat{b}^{\dagger}\hat{b}\hat{b}  +\frac{1}{2}\Lambda_{11}^{01}\,\hat{b}^{\dagger}\hat{b}^{\dagger}\hat{a}\hat{b} + \frac{1}{2}\Lambda_{00}^{01}\,\hat{a}^{\dagger}\hat{a}^{\dagger}\hat{a}\hat{b}}_{\text{Density-assisted hopping}}\\
&\quad + \underbrace{\frac{1}{4}\Lambda_{11}^{00}\,\hat{b}^{\dagger}\hat{b}^{\dagger}\hat{a}\hat{a} +\frac{1}{4}\Lambda_{00}^{11}\,\hat{a}^{\dagger}\hat{a}^{\dagger}\hat{b}\hat{b}}_{\text{Axion Pair hopping}} 
\end{aligned}
\end{equation}
where for notational simplicity, we define $0\equiv a$ and $1\equiv b$. This Hamiltonian preserves the total axion number, namely $[\hat H,\hat{n}_a+\hat n_b]=0$. The corresponding kinetic matrix elements and the gravitational quartic coefficients are given by \eqref{KineticMatrix} and \eqref{LambdaCouplings} respectively. 

\begin{figure}[t]
\centering
\includegraphics[
width=1.3\linewidth,
height=0.45\textheight,keepaspectratio
]{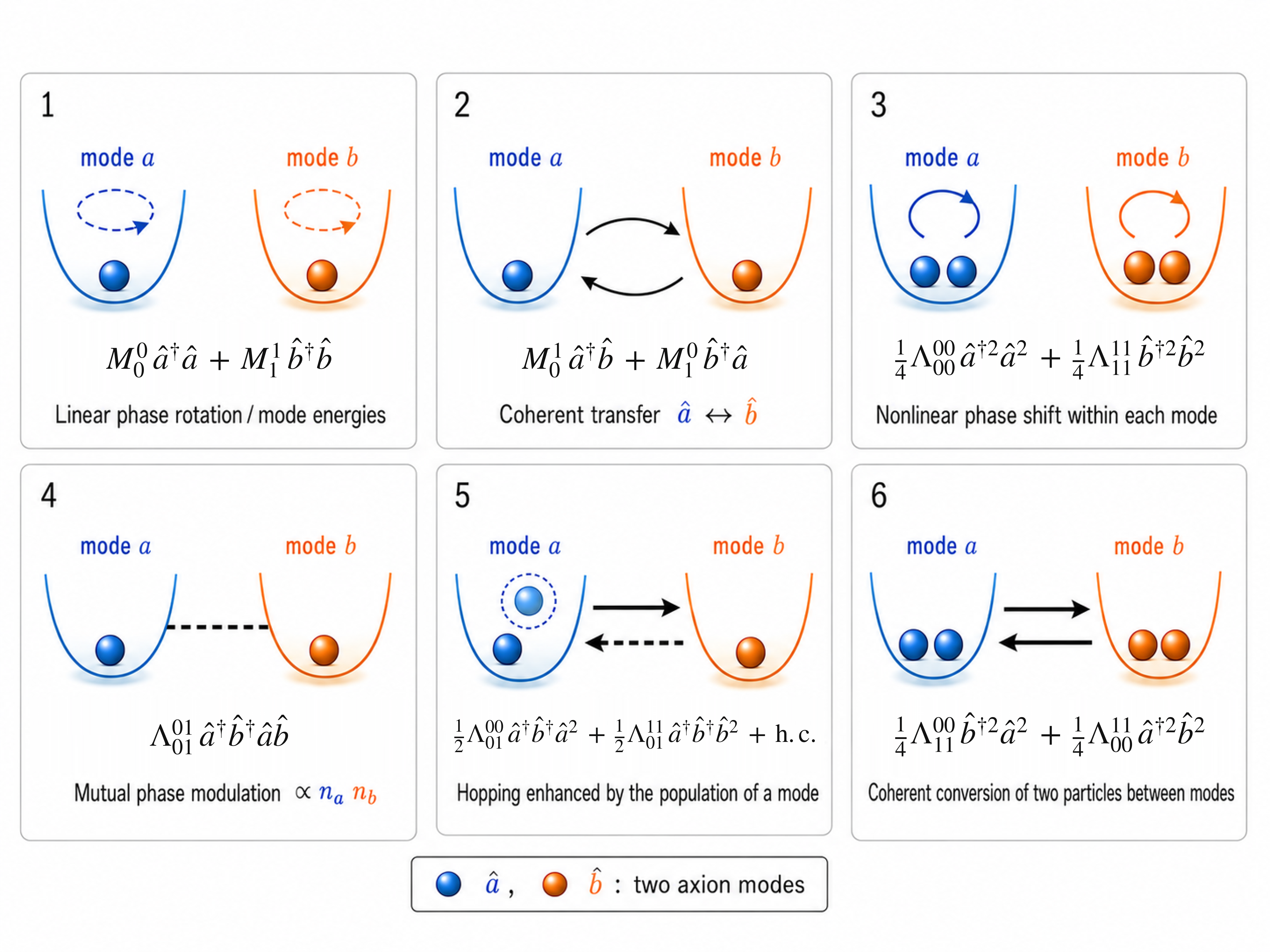}
\caption{\small Physical processes associated with the two‑mode axion Hamiltonian in Eq.\eqref{eq:twomodeHami}. For a detailed discussion of the analogous processes in quantum optics, see Ref.\cite{KockumMiranowiczMacriSavastaNori2017}. }
\label{Processes}
\end{figure}

In figure~\ref{Processes} we illustrate the physical meaning of each term in the Hamiltonian \eqref{eq:twomodeHami}. These processes are well‑studied in quantum optics \cite{KockumMiranowiczMacriSavastaNori2017}. Hamiltonian \eqref{eq:twomodeHami}, as well as simplified versions of it, as we will see below, is able to capture the quantum seeds of decoherence and the breakdown of classicality of axion dark matter from its multi-mode behavior alone -- without an external environment. However, we should mention that this Hamiltonian corresponds to a toy model in the sense that it cannot reproduce or describe effects such as the irreversibility and thermalization \cite{Erken2012,Sikivie:2009qn} that characterize a realistic many‑mode system. In that sense, it is not a fully realistic model of decoherence, but it shows that the inclusion of additional modes can reduce the squeezing effects. Our goal here is to illustrate  the relative interplay between the generation of squeezing and its loss due to mode-interactions, before considering a general open-system treatment.

We focus here on just two modes interacting through the cross-Kerr coupling, and compute the reduced dynamics of one mode after tracing out the second, showing how entanglement between the modes as well as the coupling between modes can affect the Kerr-induced squeezing.

\subsubsection{Illustrative example: Particle conservation within each mode}
To illustrate how the inter‑mode couplings can affect the squeezing, we focus on a scenario in which no particle exchange occurs between modes $a$ and $b$. That is, we keep only those terms in \eqref{eq:twomodeHami} that conserve the particle number \textit{in each mode separately}, namely $[\hat{H},\hat n_a]=[\hat{H},\hat n_b]=0$. The Hamiltonian then reads\begin{equation}\label{Ham:twomodesimple}
\hat{H} = M_{0}^{0}\,\hat{a}^{\dagger}\hat{a} + M_{1}^{1}\,\hat{b}^{\dagger}\hat{b} +  \underbrace{\frac{1}{4}\Lambda_{00}^{00}\,\hat{a}^{\dagger}\hat{a}^{\dagger}\hat{a}\hat{a}+ \frac{1}{4}\Lambda_{11}^{11}\,\hat{b}^{\dagger}\hat{b}^{\dagger}\hat{b}\hat{b}}_{\rm Self-Kerr}+\underbrace{\Lambda_{01}^{01}\,\hat{a}^{\dagger}\hat{b}^{\dagger}\hat{a}\hat{b}}_{\rm Cross-Kerr} 
\end{equation}
The first two terms describe the usual free evolution, the third and fourth terms describe the self-Kerr evolution for each mode and the last term is a cross-Kerr coupling between modes $a$ and $b$. For an initial state that is a product of coherent states, $\ket{\psi(0)} = \ket{\alpha}_a \otimes \ket{\beta}_b$, and constant $\Lambda$ over the time interval considered, the problem can be analytically solved, leading to the exact time-evolved state
\begin{equation}\label{crossKerrState}
|\psi(t)\rangle = e^{-\frac{|\alpha|^2+|\beta|^2}{2}} \sum_{n,m=0}^\infty \frac{\alpha^n\beta^m}{\sqrt{n!\,m!}}\, e^{-i\Omega_{n,m}t}\,|n,m\rangle   
\end{equation}
with angular frequency eigenvalues given by 
\begin{equation}
\Omega_{n,m}\equiv\frac{E_{n,m}}{\hbar} = M_0^0 n + M_1^1 m + \frac{1}{4}\Lambda_{00}^{00}n(n-1) + \frac{1}{4}\Lambda_{11}^{11}m(m-1) + \Lambda_{01}^{01}nm  
\end{equation}
The Fock-basis matrix elements of the density matrix $\hat{\rho}(t)=\sum_{m,n,p,q}\rho_{mn}^{pq}(t)|n,m\rangle\langle p,q|$, are
\begin{equation}
\rho_{mn}^{pq}(t)\equiv  \langle n,m|\hat{\rho}(t)|p,q\rangle = 
e^{-|\alpha|^2-|\beta|^2}\,
\frac{\alpha^n\beta^m(\alpha^*)^p(\beta^*)^q}{\sqrt{n!\,m!\,p!\,q!}}\,
e^{-i(\Omega_{n,m}-\Omega_{p,q})t}  
\end{equation}
Tracing out the $b$-mode, $\hat{\rho}_a(t)=\mathrm{Tr}_b[\hat{\rho}(t)]$, we get the reduced density of the $a$-mode
\begin{equation}\label{eq:tracetwomodes}
\hat{\rho}_a(t) = \sum_{n,p=0}^{\infty} \hat{\rho}^a_{n,p}(t)|n\rangle\langle p|=e^{-|\alpha|^2}\sum_{n,p=0}^{\infty}  \,\frac{\alpha^n(\alpha^*)^p}{\sqrt{n!\,p!}}\,
e^{-i\phi_{n,p}t}\,
\exp\!\Bigl[ |\beta|^2\Bigl( e^{-i\,\Lambda_{01}^{01}(n-p)t}-1\Bigr) \Bigr]|n\rangle\langle p|,
\end{equation}
\begin{equation}
\phi_{n,p}=M_0^0\,(n-p) + \frac{\Lambda_{00}^{00}}{4}\bigl[n(n-1)-p(p-1)\bigr] 
\end{equation}

\subsubsection{Computation of the Minimal Variance $V_-(t)$}
Having the exact solution of the reduced density matrix $\hat \rho_a(t)$ for the mode $a$, after tracing out the $b$-mode, we can compute the minimal quadrature variance
\begin{equation}
V_-(t)=1+2\left(N-|\langle \hat a(t)\rangle|^2\right)-2\left|\langle \hat a^2(t)\rangle-\langle \hat a(t)\rangle^2\right| 
\end{equation}
for the initial state $|\psi(0)\rangle=|\alpha\rangle_a\otimes|\beta\rangle_b$. A straightforward calculation of the various moments leads to the exact solution, at all times,
\begin{equation}\label{minimalvarinaceAnalytical}
\begin{split}
V_-(t)=1&+2|\alpha|^2\left[1-\exp\left(
2|\alpha|^2\left[\cos\left(\frac{\Lambda_{00}^{00} t}{2}\right)-1\right]+2|\beta|^2\left[\cos\bigl(\Lambda_{01}^{01} t\bigr)-1\right]\right)\right]
\\
&-2|\alpha|^2\Bigg|e^{-i\Lambda_{00}^{00} t/2}\exp\left[|\alpha|^2\left(e^{-i\Lambda_{00}^{00} t}-1\right)
+|\beta|^2\left(e^{-2i\Lambda_{01}^{01} t}-1\right)\right]\\
&\hspace{28mm}
-\exp\left[2|\alpha|^2\left(e^{-i\Lambda_{00}^{00} t/2}-1\right)+2|\beta|^2\left(e^{-i\Lambda_{01}^{01} t}-1\right)\right]\Bigg|
\end{split}
\end{equation}
\begin{figure}[t]
\centering
\includegraphics[width=\textwidth]
{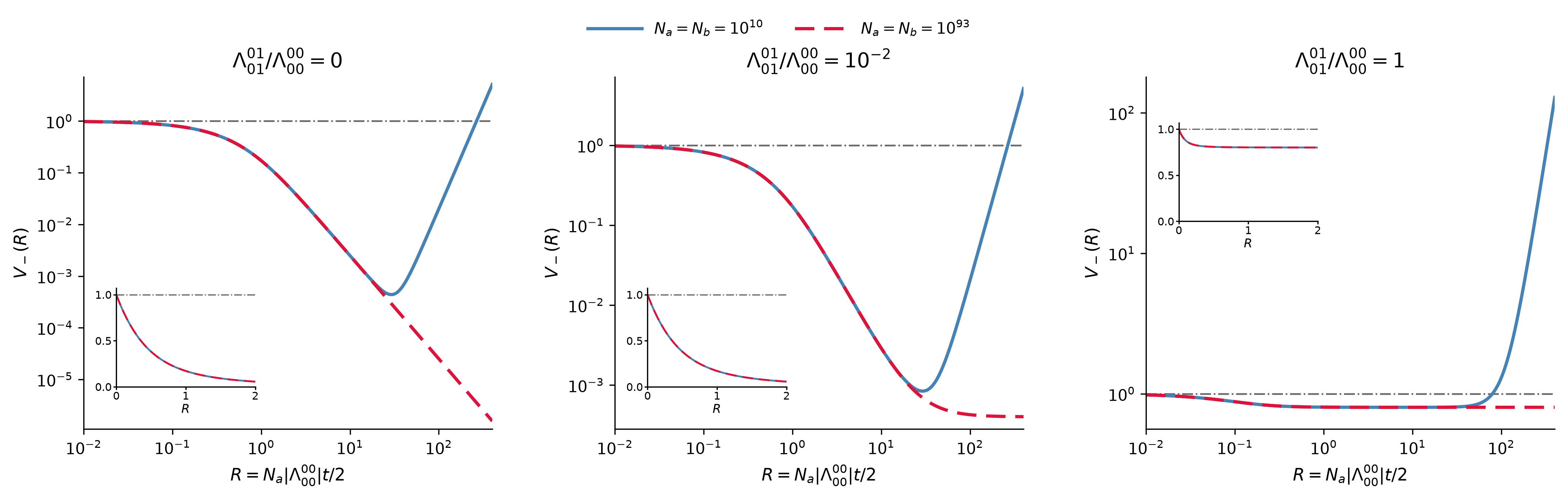}
\caption{\small Evolution of the minimal quadrature variance $V_-(t)$, as given by eq.\eqref{minimalvarinaceAnalytical}, for two equally occupied $N\equiv N_a=N_b$ modes both initially in a coherent state, comparing $N=10^{10}$ (blue solid curves) and $N=10^{93}$ (red dashed curves). From left to right, the coupling ratio of the cross to the self-Kerr couplings is $\Lambda_{01}^{01}/\Lambda_{00}^{00}=(0,10^{-2},1)$. Time is expressed through the dimensionless variable $R$. The main panels use logarithmic axes, while the insets show only the onset of evolution on linear axes. The dashed horizontal line marks the vacuum variance, $V_-=1$, below which is an indication of quadrature squeezing. Clearly, squeezing develops initially in all three cases. Increasing the inter-mode coupling reduces the achievable squeezing depth. In a similar spirit as in Appendix B in Ref.\cite{KoppFragkosPikovski2022}, using $V_{\rm min}\propto N_a^{-1/3}$, the magnitude of squeezing is $S_{\rm dB}\equiv-10\log_{10}(V_{\rm min})\simeq 34 \rm dB$ when $\Lambda_{01}^{01}=0$, $S_{\rm dB}\simeq 31$ when $\Lambda_{01}^{01}/\Lambda_{00}^{00}=10^{-2}$ and $S_{\rm dB}\simeq 1\rm dB$ when $\Lambda_{01}^{01}=\Lambda_{00}^{00}$. Note that for larger $N_a$, for instance $N_a\simeq10^{93}$ for the case of axions in a cosmological Hubble volume \cite{KoppFragkosPikovski2022,Hui2021WaveDarkMatter}, we get a deeper $V_{\rm min}$ corresponding to $S_{\rm dB}\simeq 314 \rm dB$ for the leftmost panel, as discussed in Sec.\ref{sec:2} and in Ref.\cite{KoppFragkosPikovski2022}. The middle and right panel, instead approach finite variance floors, corresponding to approximately $34\mathrm{dB}$ and $1\mathrm{dB}$, respectively. The finite squeezing and its eventual loss is here due to further unitary evolution that produces other highly non-classical states, states beyond the squeezing which is depicted in Fig. \ref{fig:squeezingKerr}, and not due to decoherence.} 
\label{twomodesqueezing}
\end{figure}
For $|\beta|^2=0$, we recover the single-mode Kerr result \cite{TanasMiranowiczKielich1991,BajerMiranowiczTanas2002}. Moreover, the self-Kerr evolution of the $b$-mode, namely terms proportional to $\Lambda_{11}^{11}$, do not appear in \eqref{minimalvarinaceAnalytical}, as expected. The reason is that the Hamiltonian of $b$-mode commutes with the cross-Kerr interaction. Since this part of the system acts only on the traced-out subsystem, it cancels under the partial trace and therefore has no effect on the reduced state of the $a$-mode.

In order to capture the onset of squeezing, at early times, it is sufficient to Taylor expand the minimal variance to get 
\begin{equation}\label{eq:Variance-two-mode-dephasing}
V_-(t)=1-|\alpha|^2|\Lambda_{00}^{00}|t+\frac{1}{2}\biggl(|\alpha|^4|\Lambda_{00}^{00}|^2+\underbrace{4|\alpha|^2|\beta|^2|\Lambda_{01}^{01}|^2}_{\rm Cross-Kerr}\biggr)t^2+\mathcal{O}(t^3)
\end{equation}
This equation makes explicit the competition between the different contributions: The linear in time self-Kerr term, $\propto -|\alpha|^2|\Lambda_{00}^{00}|t$, drives the variance below the vacuum noise level and is therefore responsible for the onset of squeezing. By contrast, the positive terms $\propto t^2$ increase the variance and therefore counteract the squeezing. The term $\propto |\alpha|^4|\Lambda_{00}^{00}|^2t^2$ corresponds to a higher-order correction to the linear in time self-Kerr term which, during the onset of squeezing, tends to drive the minimal variance back to $V_-(t)\geq 1$. The environmental correction stemming from tracing-out the $b$ mode is
\begin{equation}
\Delta V_{\rm cross}\equiv V^{(\Lambda_{01}^{01}\neq0)}_-(t)-V^{(\Lambda_{01}^{01}=0)}_-(t)=2|\alpha|^2|\beta|^2|\Lambda_{01}^{01}|^2t^2+\mathcal{O}(t^3)
\end{equation}
To gain some intuition, we define the dimensionless parameter
\begin{equation}
\epsilon\equiv 4\frac{|\beta|^2}{|\alpha|^2}\frac{|\Lambda_{01}^{01}|^2}{|\Lambda_{00}^{00}|^2}   
\end{equation}
When $\epsilon\ll 1$, that is when $|\alpha|^2|\Lambda_{00}^{00}|^2\gg |\beta|^2|\Lambda_{01}^{01}|^2$, the mode $a$ behaves almost as the usual single-mode Kerr oscillator. The coupling to the $b$-mode is not important during the onset of squeezing. On the other hand, when $\epsilon\simeq 1$, the state of $a$-mode still exhibits squeezing (recall that squeezing develops at linear order in time $t$) but the amount is reduced compared to the single-mode Kerr case, as it can be seen from Fig.\ref{twomodesqueezing}.

In other words, as long as $|\Lambda_{00}^{00}|\neq 0$, the initial slope of the minimal variance is always negative, meaning that there is always the initial squeezing.

This example highlights a crucial point for our analysis: moving from a single-mode to the two-mode axion model already captures the fact that squeezing is not automatically destroyed by the presence of an additional environmental mode $b$. The extent to which squeezing becomes substantial depends on the details of the specific model, especially the relative coupling strengths ($\Lambda_{00}^{00},\Lambda_{01}^{01}$) and the relative number of axions within each mode $N_a$ and $N_b$. 
\subsubsection{Linear entropy and entanglement between modes}
Computing the linear entropy $S_L(t) = 1 - \operatorname{Tr}\!\bigl(\hat{\rho}_a^2(t)\bigr)$, we get
\begin{equation}
 S_L(t) = 1 - e^{-2|\alpha|^2} \sum_{n=0}^{\infty}\sum_{p=0}^{\infty}
\frac{|\alpha|^{2n}}{n!}\frac{|\alpha|^{2p}}{p!}\,
\exp\!\Bigl[ -2|\beta|^2\Bigl(1 - \cos\!\bigl[\Lambda_{01}^{01}\,(n-p)\,t\,\bigr]\Bigr) \Bigr]   \, .
\end{equation}
The entropy captures the entanglement generated between the two modes, for a bipartite pure state.
We can immediately observe that at initial time $t=0$, the linear entropy vanishes, $S_L(0)=0$, as expected for a product state. On the other hand, at any time $t>0$, the linear entropy is positive $S_{L}>0$, except for the times $t\equiv\tau_{rev}=\frac{2\pi k}{\Lambda_{01}^{01}}$, for $k\in Z$, that are the so-called revival times and $S_L(\tau_{rev})=0$ holds. A positive linear entropy therefore establishes entanglement between the modes $a$ and $b$.

This simple calculation illustrates that, under the cross-Kerr coupling, an initial pure product state of coherent states evolves into an entangled two-mode state. Tracing out the $b$-mode, leads to a mixed state for the $a$-mode, which \textit{cannot be described by a single coherent state}. For axion dark matter, such couplings are inevitably present, even in absence of any external environment, since gravitational self‑interactions generate both self‑Kerr and cross‑Kerr terms together with mode-changing interaction terms, as is clear from eq.\eqref{eq:twomodeHami}.  

We should keep in mind that the “environment” in the previous example consists only of a single mode ($b$), and that the decoherence is periodic and reversible. At certain times the off‑diagonal factor in \eqref{eq:tracetwomodes} returns to 1 and the coherence revives completely. In other words, this is not an irreversible\footnote{It should be noted that all physical decoherence has in principle a revival, as the global evolution is unitary. The irreversibility expected for decoherence is thus only effective, and depends on the physical situation.} Lindblad‑type decoherence but simply a unitary dephasing and re-phasing cycle, similar in spirit to the Jaynes-Cummings model in Quantum optics \cite{JaynesCummings1963}. In section \ref{sec:Decoherence of ADM}, we will discuss the phenomenological approach to axion dark matter decoherence using Markovian Lindblad equations. In particular, we return to the case of dephasing for a generic bath in section \ref{sec:Dephasing}.

\subsection{Transition to a generic quantum open system analysis: Markovian master equation in Lindblad form}\label{sec:bridgingtwomodeLindblad}

The previous two-mode example illustrates how coupling to an unobserved degree of freedom generates entanglement and modifies the squeezing of a selected axion mode. However, a more realistic scenario will involve other couplings, additional unresolved axion modes or other environmental degrees of freedom. To investigate the robustness of Kerr squeezing across such settings, in Section \ref{sec:Decoherence of ADM}, we use an effective description that captures the relevant environmental processes without requiring a complete microscopic model of axions and each environment separately. The Lindblad framework provides such a description: If the reduced dynamics of the selected system is Markovian, then, the generator of completely positive and trace preserving (CPTP) dynamics \textit{must} have the Gorini-Kossakowski-Sudarshan-Lindblad (GKLS) form  \cite{Lindblad1976,Gorini1976,Kossakowski1972}

\begin{equation} \label{eq:Lindblad}
\frac{d}{dt}\hat{\rho}_a = -i[\hat H_a , \hat \rho_a]
+ \sum_{\mu} \gamma_{\mu}
\biggl( \hat S_{\mu}\hat\rho_a \hat S_{\mu}^\dagger
- \frac{1}{2} \left\{\hat S_{\mu}^\dagger\hat S_{\mu},\hat \rho_a\right\} \biggr)  
\end{equation}
where $\hat \rho_a$ denotes the reduced density matrix of the selected system (for instance the $a$-mode of the axion field), $\hat H_a$ is the system Hamiltonian and $\left\{,\right\}$ is the anti-commutator. Once the operators $\hat{S}$ have been specified, all the remaining influence of the environment is encoded in the rates $\gamma\geq 0$ while the dissipator $\mathcal{D}[\hat{S}]\hat \rho_a\equiv \hat S\hat\rho_a \hat S^\dagger-\frac{1}{2} \left\{\hat S^\dagger\hat S,\hat \rho_a\right\}$ captures the non-unitary piece of the dynamics \cite{BreuerPetruccione2002}.

This equation provides a very \textit{general} description of a selected quantum system interacting with unresolved environmental degrees of freedom, in the Markovian limit.  Its strength is precisely its generality: this equation captures a very broad class of physical environments whose detailed microscopic origin (their exact Hamiltonian dynamics) may significantly differ and lead to different jump operators $\hat{S}$ but, nevertheless, lead to the same Lindblad structure, Eq.\ref{eq:Lindblad}. The different physical scenarios are captured by phenomenologically choosing different jump operators.

The inter-mode couplings in the axion Hamiltonian provide microscopic motivation for several possible environmental channels. Couplings that exchange particles between the selected mode and unresolved modes motivate loss and gain processes, while occupation-dependent couplings, such as Cross-Kerr couplings, motivate phase noise.

In Section 4, we study loss, pure dephasing, and finite-temperature damping phenomenologically and determine their effects on the onset, magnitude, and duration of Kerr squeezing. The resulting conditions apply to different physical environments whenever the corresponding effective channel is justified: Application to a particular axion setting requires estimates of the relevant environmental parameters.
\section{Effect of Generic Environments on Kerr Squeezing}

\label{sec:Decoherence of ADM}
The discussion so far was based on closed, unitary dynamics, where one could integrate out parts of the system analytically. In Section 2, we modeled the axion field as a single macroscopically occupied mode $\hat{\psi}(\boldsymbol{x},t)\simeq\frac{1}{\sqrt{N}}\psi(\boldsymbol{x},t)\hat{a}(t)$ whose gravitational self-interaction includes a Kerr nonlinearity term and hence leads to quadrature squeezing. In Section 3, we showed that a two-mode extension model, $\hat{\psi}(\boldsymbol{x},t)\simeq u_1(\boldsymbol{x},t)\hat{a}(t)+u_2(\boldsymbol{x},t)\hat{b}(t)$, produces entanglement between modes, which, after tracing over one of them, affects the dynamics of squeezing.

However, a realistic axion field will interact with a larger environment. A selected axion mode may interact with unresolved axion modes and, depending on the model and the environment, with photons, matter or any other coarse-grained environmental degrees of freedom. A microscopic treatment of these interactions would require specifying the relevant environmental degrees of freedom, their spectral properties as well as their couplings to the selected axion mode. These ingredients vary from one physical setting to another and therefore such a treatment would entail highly environmental-specific calculations rather than a single general description, and in many cases are intractable analytically. Therefore, instead of attempting to model various possibilities microscopically, in this section, we adopt an \textit{open-system perspective} to identify how \textit{robust} the Kerr squeezing mechanism is against generic environmental decoherence channels.

Here, we keep the same system Hamiltonian $\hat H\equiv\hat{H}_{\rm phys}/\hbar=\omega \hat a^\dagger \hat a+\chi \hat a^{\dagger 2}\hat a^2$ as in \cite{KoppFragkosPikovski2022} but allow the reduced density matrix of the selected mode to evolve according to a Markovian master equation in Lindblad form, $\dot\rho=-i[\hat H,\rho]+\mathcal L_{\rm env}[\rho]$ \cite{Lindblad1976,Gorini1976,BreuerPetruccione2002}.

Although the precise environmental rates will differ for each microscopic process, a small set of Lindblad channels is capable of capturing a broad class of physical situations. We will focus on three specific Lindblad channels: Zero temperature damping, pure dephasing channel and finally, damping by a finite-temperature bath of simple harmonic oscillators. 

Zero-temperature damping models the irreversible loss of axions from the selected axion mode to the environment. Pure dephasing describes phase randomization without any particle loss, similar in spirit to the cross-Kerr effect previously studied in \ref{Subsection:2-mode case}. Finally, finite-temperature damping describes the competition between loss, absorption, and thermal noise \cite{WallsMilburn2008}.

For each case, we compute the minimal variance $V_-(t)$ and determine the conditions under which it falls below the vacuum level. In the unitary problem, under the initial coherent state assumption, once the self-Kerr non-linearity is instantiated, the minimal variance immediately drops below the vacuum level and hence squeezing begins at arbitrarily small positive times, while the characteristic time to reach $r(t_{\rm sqz})=1$ is $t_{\rm sqz}\simeq0.6/N|\chi|$, as defined in \eqref{unitarytsqz}. Our aim is to establish how each environmental channel modifies the evolution and, in particular, under which conditions the onset of squeezing remains intact and appreciable squeezing survives.

Throughout the following sections, unless otherwise stated, we denote by $N\equiv |\alpha|^2$ the initial mean occupation of the selected axion mode. Since dissipative channels do not necessarily conserve the number of particles in this mode, the instantaneous occupation, when required, will be denoted explicitly by $N(t)\equiv \langle \hat a^\dagger \hat a\rangle_t $.
\subsection{Dissipation of Axions to a Zero-Temperature Environment}
\label{sec:zeroTemp}
As a first open-system extension of the unitary Kerr dynamics, we consider the simplest irreversible process: loss of axions from the selected axion mode into an unobserved environment at zero temperature.  In the Markovian limit, this loss is described by the standard amplitude-damping Lindblad equation \cite{BreuerPetruccione2002,RossiAlbarelliParis2016} 
\begin{equation}\label{zerotempmaster}
\frac{d}{dt}\hat{\rho}=-i[\hat{H},\hat{\rho}]+\gamma\biggl(\hat{a}\hat{\rho}\hat{a}^{\dagger}-\frac{1}{2}\hat{a}^{\dagger}\hat{a}\hat{\rho}-\frac{1}{2}\hat{\rho}\hat{a}^{\dagger}\hat{a}\biggr)
\end{equation}
Here, $\hat \rho$ is the reduced density matrix of the selected axion mode, obtained by tracing over the environment. $\hat{H}=\omega \hat a^\dagger \hat a+\chi \hat a^{\dagger 2}\hat a^2$ is the self-Kerr Hamiltonian \eqref{kerrHamiltonian} and $\gamma$ is the axion loss rate. The corresponding jump operator in the Lindblad equation is $\hat L=\sqrt{\gamma}\hat a$, so each jump removes one axion from the selected mode. Since the environment is at zero temperature, there is no inverse process proportional to $\hat a^\dagger$. In other words, the environment absorbs quanta from the axion mode but does not inject them back.

Equation \eqref{zerotempmaster} can be analytically solved, for an initial coherent state $\hat{\rho}_0=|\alpha\rangle\langle\alpha|$, to obtain \cite{miranowicz1990generation,RossiAlbarelliParis2016,Fragkos2025Thesis}
\begin{equation}\label{Parissolution}
\hat{\rho}(t)=e^{-|\alpha|^2}\sum_{m,n=0}^{\infty} \frac{\bigl(\alpha e^{-\frac{\gamma t}{2}}e^{-i(\omega-\chi)t}\bigr)^m}{\sqrt{m!}}\frac{\bigl(\alpha^* e^{-\frac{\gamma t}{2}}e^{+i(\omega-\chi)t}\bigr)^n}{\sqrt{n!}}e^{-i\chi t(m^2-n^2)}e^{A_{mn}(\chi,\gamma)}|m\rangle \langle n|
\end{equation}
where 
\begin{equation}\label{coefficientParis}
A_{mn}(\chi,\gamma)=\frac{\gamma |\alpha|^2}{\gamma+2i\chi(m-n)}\biggl(1-e^{-\bigl(\gamma+2i\chi(m-n)\bigr)t}\biggr)
\end{equation}
The function $A_{mn}(\chi,\gamma)$ encodes the interplay between particle loss and self-Kerr evolution. In the absence of losses, that is when $\gamma=0$, we get $A_{mn}(\chi,\gamma=0)=0$ and Eq. \eqref{Parissolution} reduces to the pure-state unitary self-Kerr solution \cite{YurkeStoler1986,Fragkos2025Thesis}. Conversely, when self-Kerr evolution is absent, i.e when $\chi=0$, then $A_{mn}(\chi=0,\gamma)=N(1-e^{-\gamma t})$ is independent of the number-state indices $(n,m)$ and the initial coherent state remains coherent, namely $|\alpha e^{-\gamma t/2-i\omega t}\rangle$ \cite{WallsMilburn2008,WallsMilburn1985DissipationCoherence}.

When both processes are present, $A_{mn}(\chi,\gamma)$ acquires a dependence on $(m-n)$. Loss events change the occupation during the nonlinear phase evolution, and averaging over unobserved loss generally reduces coherence between number states. However, it is crucial to point out that this loss of purity \textit{does not} by itself determine whether quadrature squeezing survives. We therefore turn to the minimal quadrature variance to quantify how amplitude damping modifies the generation of Kerr squeezing.
\subsubsection{Computation of Minimal Variance for zero-temperature environment}
We now quantify the effect of zero-temperature damping on the unitary Kerr squeezing through the minimal quadrature variance $V_-(t)$. For an initial coherent state $|\alpha\rangle$, the relevant moments for computing the minimal variance \eqref{eq:minimalvarianceDefinition} are
\begin{equation}
\langle \hat{a}^{q}\rangle_t=\alpha^q e^{-q\gamma t/2}e^{-iq\omega t}\mathcal{E}_q(t),\qquad  \langle \hat{a}^{\dagger}\hat{a} \rangle_t=Ne^{-\gamma t}
\end{equation}
where
\begin{equation}\label{varzerotemp}
\mathcal{E}_{q}(t)\equiv\exp\left[-iq(q-1)\chi t
+N\frac{2iq\chi}{\gamma+2iq\chi}\left(e^{-(\gamma+2iq\chi)t}-1\right)\right]
\end{equation}
for $q=1,2$.
The result for the minimal variance is \cite{Fragkos2025Thesis}
\begin{equation}\label{eq:MinimalvarianceT=0atalltimes}
V_-(t)=1+2Ne^{-\gamma t}\biggl[1-|\mathcal{E}_1(t)|^2-|\mathcal{E}_2(t)-\mathcal{E}^2_1(t)| \biggr]
\end{equation}
The solution to the Lindblad equation as given by eq.\eqref{Parissolution} and the corresponding minimal variance \eqref{eq:MinimalvarianceT=0atalltimes} are \textit{exact} solutions which hold at all times $t\in[0,\infty)$.
\begin{figure}[t]
\centering
\includegraphics[width=\textwidth]
{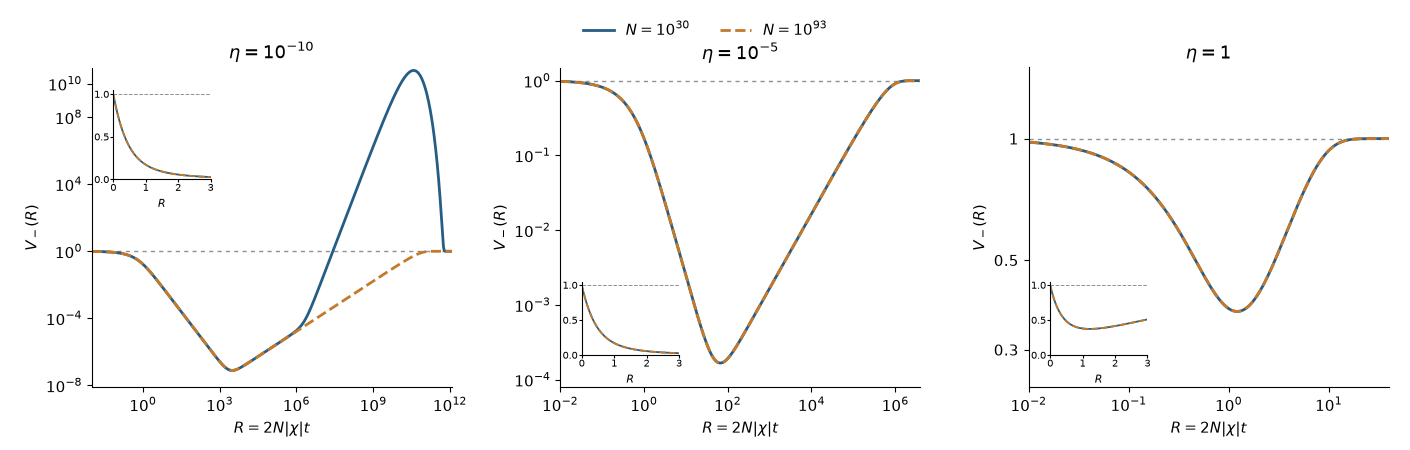}
\caption{\small  Evolution of minimal quadrature variance, Eq. (\ref{eq:MinimalvarianceT=0atalltimes}), under Kerr dynamics with zero-temperature damping, for two different occupation numbers $N$. From left to right, the damping parameter is $\eta\equiv\gamma/(N|\chi|)=(10^{-10},10^{-5},1)$. Squeezing develops in all three cases, with stronger damping $\eta$ reducing significantly its depth. For the weakest damping, $\eta=10^{-10}$, the occupations differ at late times, with a pronounced rise above the vacuum level for $N=10^{30}$. All curves eventually approach the vacuum variance as the mode decays. The insets illustrate the onset of squeezing at timescales $R\sim\mathcal{O}(1)$. For $\eta\ll1$, the curves are effectively described by the unitary result, as shown analytically by Eqs. \eqref{eq:splittingzerotemp} and \eqref{eq:T=0dissipationcorrection}. For mathematical details on these scalings the reader is referred to Appendix \ref{app:zero_Temp_tmax}.}
\label{fig:zeroTemp}
\end{figure}
\subsubsection{Finite-$R$ approximation for the zero-temperature environment}
Although eq.\eqref{eq:MinimalvarianceT=0atalltimes} gives the exact variance at all times, for the highly occupied modes relevant to axion dark matter, appreciable Kerr squeezing develops when $N|\chi|t=\mathcal{O}(1)$ even though $|\chi|t\ll1$ \cite{KoppFragkosPikovski2022}. We therefore expand the exact expression for $N\gg 1$ while retaining the full dependence on $R\equiv2N|\chi|t$, to end up with
\begin{equation}\label{minimalvardissip}
\boxed{V_-(t)
\simeq 1+2R^2-2R\sqrt{1+R^2}+\frac{1}{N}\left[\frac{R^3(5+3R^2)}{\sqrt{1+R^2}}
-R^4+\underbrace{\frac{\gamma}{3|\chi|}\left(\frac{3R^2+5R^4}{\sqrt{1+R^2}}-5R^3\right)}_{\mathrm{Environment}}\right]+\mathcal{O}\bigl(N^{-2}\bigr)}
\end{equation}
where
\begin{equation}
R \equiv N\tau = |\alpha|^2 \tau,\qquad 
\tau \equiv 2|\chi| t,\qquad N \equiv |\alpha|^2 
\end{equation}
One can intuitively think of the parameter $R$ as a "clock" that measures how much the collective non-linear evolution that creates squeezing has progressed.

Equation \eqref{minimalvardissip} is obtained as follows: We introduced the parameter $R \equiv N\tau = |\alpha|^2 \tau$ where $\tau\equiv2|\chi| t$ and expanded the minimal variance, as given by eq.\eqref{eq:MinimalvarianceT=0atalltimes}, for $N\gg 1$ and $\tau\ll1$, keeping $R$ and $\gamma/|\chi|$ fixed, retaining the leading term and the first correction in $1/N$ \cite{KoppFragkosPikovski2022, BajerMiranowiczTanas2002,Fragkos2025Thesis}. 

For $\gamma=0$, we recover eq. B3 in \cite{KoppFragkosPikovski2022}. To gain further intuition, we now expand Eq.~\eqref{minimalvardissip} in the small-$R$ regime, $R\ll 1$. This is an additional approximation: in deriving Eq.~\eqref{minimalvardissip}, we kept $R\equiv N\tau=2N|\chi| t$ finite, whereas here we further restrict to the earliest part of the evolution. 
Although this limit does not describe the full finite-$R$ squeezing curve, it is sufficient to capture the initial onset of squeezing. We find
\begin{equation}
V_-(t)\simeq 1+2R^2-2R+\frac{\gamma}{N|\chi|}R^2+\mathcal{O}\biggl(R^3,\frac{R^3}{N}\biggr)
\end{equation}
or equivalently\footnote{Note that the approximation \eqref{Vminusdissip} can also be obtained without a large-$N$ approximation. One could directly Taylor expand equation \eqref{varzerotemp} at short times while imposing $\gamma t\ll 1$, $\chi t\ll 1$ and $N\chi t\ll1$. These conditions constrain the time interval of the approximation rather than requiring large occupation $N\gg 1$. }
\begin{equation}\label{Vminusdissip}
V_-(t)\simeq1-4N|\chi|t+\biggl(8N^2\chi^2+\underbrace{4\gamma N|\chi|}_{\mathrm{dissipation}}\biggr)t^2+\mathcal O(t^3)
\end{equation}
The correction due to dissipation, at $t\rightarrow 0^+$, is therefore
\begin{equation}\label{gammaonsetcorrection}
\Delta_\gamma V_-(t)=4\gamma N|\chi|t^2 +\mathcal{O}(t^3)
\end{equation}
This term, coming with a positive sign, makes $V_-(t)$ larger. However, notice that it appears as a subdominant term in the $t$-expansion. Starting from a coherent state, that is $V_-(t=0)=1$, equation \eqref{Vminusdissip} implies that $\frac{dV_-}{dt}|_{t=0^+}=-4N|\chi|<0$ meaning that the curve starts moving downward from the vacuum level. Therefore, for sufficiently small $t$, squeezing occurs. In other words, coupling of the system to a zero-temperature environment does not prevent the initial onset of squeezing, which follows the unitary result (see eq. B3 in Ref.\cite{KoppFragkosPikovski2022}). 

This has a simple physical interpretation. A zero-temperature loss channel alone maps a coherent state into another coherent state, and therefore leaves the quadrature variance at the vacuum level, i.e. $V_-(t)=1$. Its first effect on the squeezing appears only after the self-Kerr interaction has started building up correlations, which is why the dissipative correction $\Delta_{\gamma}V_-(t)$ enters at order $\sim\gamma N|\chi|t^2$ \cite{WallsMilburn2008}. 

Having established that the zero-temperature damping \textit{does not a priori prevent the initial onset of squeezing}, we are now asking a stronger question: Does the damping to a zero-temperature environment remain negligible over the timescale that the system develops $r>\mathcal{O}(1)$ squeezing? 

To determine whether appreciable squeezing develops, we return to our central result \eqref{minimalvardissip}. It is helpful to rewrite it as 
\begin{equation}\label{eq:splittingzerotemp}
 V_-(t)\simeq V^{\gamma=0}_-(t)+  \Delta_{\gamma}V_-(t) 
\end{equation}
with $V^{\gamma=0}_-(t)\simeq 1+2R^2-2R\sqrt{1+R^2}+\frac{1}{N}\bigl(\frac{R^3(5+3R^2)}{\sqrt{1+R^2}}
-R^4\bigr)$ while the leading environmental contribution is 
\begin{equation}\label{eq:T=0dissipationcorrection}
\Delta_{\gamma}V_-(t)\simeq\frac{\gamma}{N|\chi|}\frac{1}{3}\left(\frac{3R^2+5R^4}{\sqrt{1+R^2}}-5R^3\right)
\end{equation}
Before proceeding, it is worth highlighting the crucial feature of Eq. \eqref{minimalvardissip}, namely its applicability for $R\sim{O}(1)$. That is precisely the regime where the unitary variance has fallen substantially below the vacuum level, as also illustrated in Fig.\ref{fig:zeroTemp}. On the other hand, equation \eqref{Vminusdissip}, which requires $R\ll 1$, cannot capture that regime. Equation \eqref{eq:T=0dissipationcorrection} reveals that the damping is parametrically controlled by the ratio 
\begin{equation}
 \eta\equiv\frac{\gamma}{N|\chi|}   
\end{equation}
For $\eta\ll1\Longleftrightarrow\gamma \ll N|\chi|$, damping can be treated as a small perturbation during the formation of Kerr squeezing. Consequently, zero-temperature amplitude damping does not prevent the onset of Kerr squeezing and, crucially, appreciable squeezing remains close to the unitary prediction. If instead $\eta\simeq1\Longleftrightarrow\gamma \simeq N|\chi|$, squeezing still forms, but its depth is reduced as compared to the unitary estimate. This can be seen analytically from Eq.(\ref{eq:MinimalvarianceT=0atalltimes}) and Fig.\ref{fig:zeroTemp}. In this latter case, Eq.\eqref{minimalvardissip} is insufficient to capture the $\eta=\mathcal{O}(1)$ regime. 

At later times, $R\gg1$, weak damping can modify the maximum squeezing. In this case, under the condition $N^{-1/2}\ll\eta\ll1$, we get $R_{\rm max}^{(\gamma)}\simeq (3/\eta)^{1/3}$ and $V_{\rm min}^{(\gamma)}\simeq (3^{1/3}/4)\eta^{2/3}\ll1$. Therefore, minimum variance gets higher and is reached earlier than in the unitary evolution \cite{KoppFragkosPikovski2022}. Nevertheless, substantial squeezing survives. The derivation of these results is presented in Appendix \ref{app:zero_Temp_tmax}.

\subsection{Effect of a dephasing channel on Kerr squeezing} \label{sec:Dephasing}
We now consider the same unitary self-Kerr dynamics as before, but this time coupled to a dephasing environment. The Lindblad equation in this case is \cite{WallsMilburn2008}
\begin{equation}\label{eq:Lindblad.dephasing}
\frac{d}{dt}\hat{\rho}=-i\biggr[\omega \hat n+\chi \hat a^{\dagger 2}\hat a^2,\hat{\rho}\biggl]+\gamma_{\mathrm{deph}}\left(\hat n \hat{\rho} \hat n-\frac{1}{2}\left\{\hat n^2,\hat{\rho}\right\}\right)
\end{equation}
where $\gamma_{\mathrm{deph}}$ is the dephasing rate. This channel, as also discussed in Sec.\ref{Subsection:2-mode case}, preserves the particle number, that is $[\hat H,\hat n]=0$ and $[\hat n,\hat{L}(\hat{n}) ]=0$, where $\hat{L}(\hat{n})\equiv\sqrt{\gamma_{\rm deph}}\hat n$,  the Lindblad jump operator. For an initial coherent state $\hat{\rho}(0)=\ket{\alpha}\langle\alpha|$, the \textit{exact} solution to Eq.\eqref{eq:Lindblad.dephasing} is 
\begin{equation}
\hat \rho(t)=e^{-|\alpha|^2}\sum_{m,n=0}^{\infty}\frac{(\alpha e^{-i(\omega-\chi)t})^m}{\sqrt{m!}}\frac{(\alpha^*e^{+i(\omega-\chi)t})^n}{\sqrt{n!}}e^{-i\chi(m^2-n^2)t}e^{-\frac{\gamma_{\rm deph}}{2}(m-n)^2t}|m\rangle\langle n|
\end{equation}
In absence of self-Kerr squeezing, that is setting $\chi=0$, we recover the known results \cite{WallsMilburn1985DissipationCoherence,WallsColletMilburnDephasing}. Having the exact expression of the density matrix, we can now compute various moments using $\langle \hat{\mathcal{O}}\rangle\equiv  Tr[\hat \rho \hat{\mathcal{O}}]$. For the minimal quadrature variance 
\begin{equation}\label{minimalvARIANCEdefinition}
V_-(t)=1+2\left(N-|\langle \hat a(t)\rangle|^2\right)-2\left|\langle \hat a^2(t)\rangle-\langle \hat a(t)\rangle^2\right| 
\end{equation}
the relevant moments are
\begin{equation}
\langle \hat n(t)\rangle = N 
\end{equation}
\begin{equation}
\langle \hat a(t)\rangle=\alpha e^{-i\omega t}
e^{-\gamma_{\mathrm{deph}}t/2}\exp\left[N\left(e^{-2i\chi t}-1\right)\right]
\end{equation}
and
\begin{equation}
\langle \hat a^2(t)\rangle=\alpha^2 e^{-2i(\omega+\chi)t}
e^{-2\gamma_{\mathrm{deph}}t}\exp\left[N\left(e^{-4i\chi t}-1\right)
\right].
\end{equation}
Expanding the moments at short times, including terms up to $\mathcal{O}(t^2)$, we end up with
\begin{align}\label{eqVminusDephasing}
V_-(t)\simeq1&-2N\left(\mathcal{R}_{\mathrm{deph}}-\gamma_{\mathrm{deph}}\right)t+\notag\\&+\left[8N^2\chi^2\biggl(1-\frac{\gamma_{\mathrm{deph}}}{\mathcal{R}_{\mathrm{deph}}}\biggr)+3N\gamma_{\mathrm{deph}}\biggl(\mathcal{R}_{\mathrm{deph}}-\frac{1}{3}\gamma_{\mathrm{deph}}\biggr)\right]t^2+O(t^3)
\end{align}
where
\begin{equation}
\mathcal{R}_{\mathrm{deph}}\equiv\sqrt{\gamma_{\mathrm{deph}}^2+4\chi^2}
\end{equation}
In the limit  $N\gg 1$, while keeping $\gamma_{\rm deph}/\chi$  fixed, and keeping only the leading terms as in \cite{KoppFragkosPikovski2022}, the minimal variance becomes
\begin{equation}\label{minimalVDephasing}
\boxed{
V_-(t)\simeq1-2N\left(\sqrt{\gamma_{\mathrm{deph}}^2+4\chi^2}-\gamma_{\mathrm{deph}}\right)t+8N^2\chi^2
\left(1-\frac{\gamma_{\mathrm{deph}}}{\sqrt{\gamma_{\mathrm{deph}}^2+4\chi^2}}\right)t^2+\mathcal{O}(t^3)}
\end{equation}
\begin{figure}[t]
\centering
\includegraphics[width=\textwidth]
{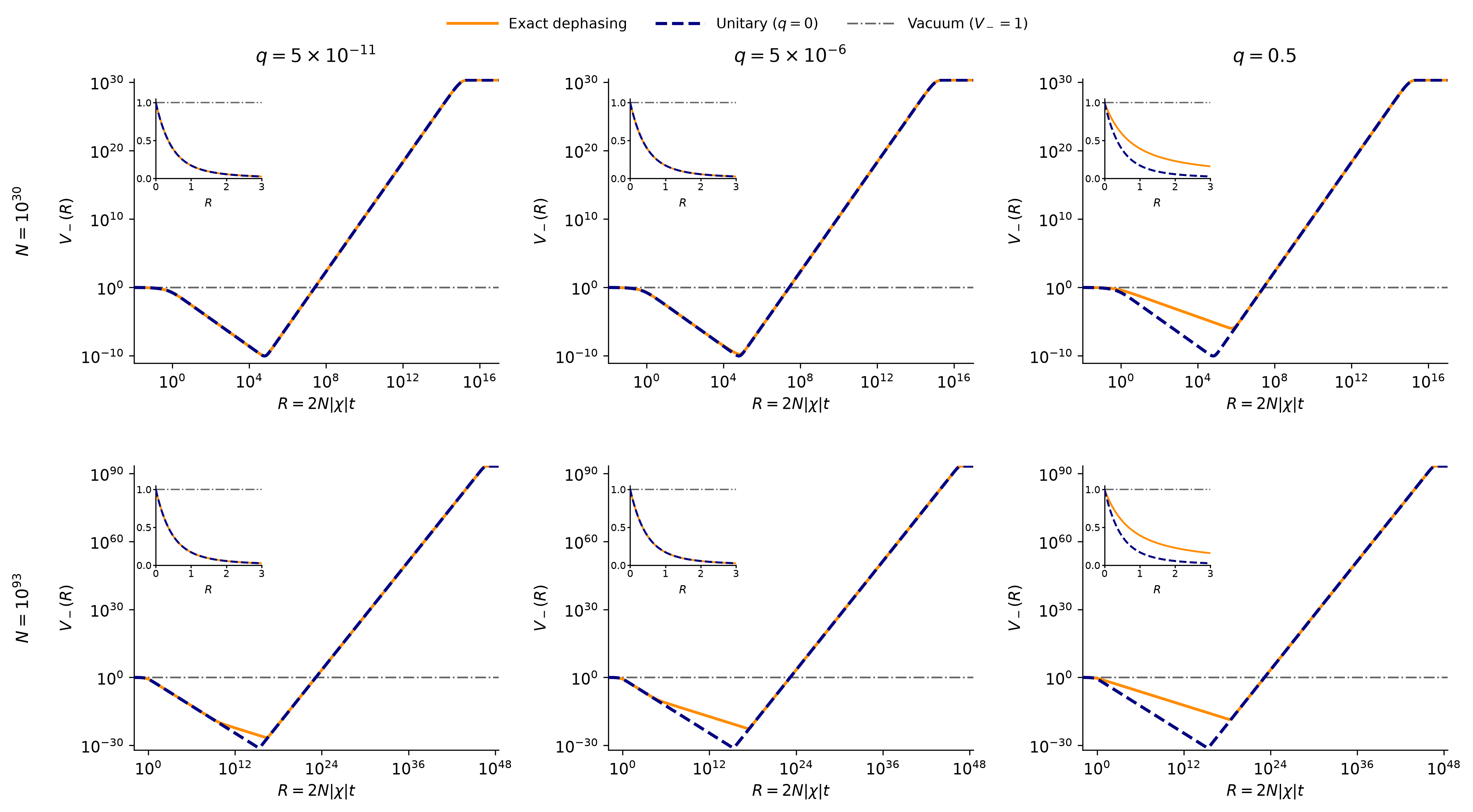}
\caption{\small Minimal quadrature variance $V_-(t)$ under Kerr evolution with pure dephasing, evaluated from the exact moments using Eq.\eqref{minimalvARIANCEdefinition}, plotted against $R=2N|\chi|t$. The upper and lower rows correspond to $N=10^{30}$ and $N=10^{93}$, respectively. The latter is the cosmological Hubble-volume benchmark for the representative QCD axion mass $m_a\simeq10^{-5}\rm eV/c^2$, considered in \cite{KoppFragkosPikovski2022}. From left to right, the dephasing strength is $q\equiv\gamma_{\mathrm{deph}}/(2|\chi|)
=(5\times10^{-11},\,5\times10^{-6},\,1/2)$. Dashed blue curves correspond to the unitary result ($q=0$) whereas horizontal dotted lines mark $V_-(R)=1$. Orange lines in the rightmost panels illustrate that stronger dephasing raises the minimum variance and delays it to later times. The insets illustrate the onset of squeezing at $R=\mathcal{O}(1)$. In all cases, the minimal variance falls below its vacuum value. The location and the value of the minimum variance and further mathematical details as well, are given in the Appendix \ref{app:dephasing}.}.
\label{fig:DephasingLindblad}
\end{figure}
In the limit $\gamma_{\mathrm{deph}}=0$, we recover the short-time/small $R\equiv N\tau\ll1$ of the unitary Kerr result, eq. B3 in \cite{KoppFragkosPikovski2022}. 
For weak dephasing, $\gamma_{\mathrm{deph}}\ll |\chi|$ we have $\mathcal{R}_{\mathrm{deph}}\simeq2|\chi|+\mathcal{O}\bigl(\frac{\gamma_{\mathrm{deph}}^2}{|\chi|}\bigr)$ and therefore
\begin{equation}
V_-(t)=1-4N|\chi|t+2N\gamma_{\mathrm{deph}}t+\mathcal{O}\biggl(\frac{N\gamma_{\rm deph}^2}{|\chi|}t\biggr)+\mathcal{O}(t^2)
\end{equation}
This expression makes the physical process clear: The Kerr term, responsible for the squeezing generation, is $V_-^{\rm Kerr}\sim-4N|\chi|t$ while dephasing gives $V_-^{\mathrm{deph}}\sim+2N\gamma_{\mathrm{deph}} t$. Thus, unlike zero-temperature damping case previously studied, pure dephasing kicks in at the same order in time as the self-Kerr squeezing itself. Already at order $t$, while self-Kerr dynamics try to reduce the minimal variance $V_-(t)$ below the vacuum noise, $V_-(t)<1$, dephasing suppresses the phase coherence needed in order to build quadrature squeezing and tends to increase $V_-$ relative to the unitary prediction.

Consequently, from equation \eqref{minimalVDephasing}, we see that the \textit{initial slope} of the minimal variance
\begin{equation}\label{slopedephasing}
\frac{dV_-}{dt}|_{t=0^+}=-2N\left(\sqrt{\gamma_{\mathrm{deph}}^2+4\chi^2}-\gamma_{\mathrm{deph}}\right)<0
\end{equation}
is negative for any value of $\gamma_{\rm deph}>0$. Therefore, dephasing \textit{does not} a priori remove the mathematical onset of squeezing but it can reduce the initial rate at which squeezing develops, depending on the relation between $\gamma_{\mathrm{deph}}$ and $\chi$.  On the other hand, in the strong-dephasing regime, $\gamma_{\mathrm{deph}}\gg |\chi|$,\footnote{The limit $\gamma_\mathrm{deph}\gg |\chi|$ does not by itself invalidate the phenomenological Lindblad model. The environmental rate $\gamma_{\rm deph}$ has to be small compared with the inverse bath-correlation time, $\gamma_\mathrm{deph}\tau_B\ll 1$, but not necessarily small compared with the Kerr coefficient $|\chi|$. The short-time expansion is trustworthy provided, in addition, $\gamma_\mathrm{deph} t\ll 1$ and $N|\chi|t\ll 1$.} the minimal variance becomes\footnote{To derive this expression, it is important to mention that apart from $\gamma_{\rm deph}\gg|\chi|$, the condition $\frac{\gamma_{\rm deph}}{|\chi|}\ll N^{1/4}$ needs to be satisfied. This can be seen from the expression in the square bracket of eq. \eqref{eqVminusDephasing} that multiplies $t^2$.}
\begin{equation}
V_-(t)\simeq1-\frac{4N\chi^2}{\gamma_{\rm deph}}\,t+\frac{16N^2\chi^4}{\gamma_{\rm deph}^2}\,t^2+\mathcal{O}(t^3)
\end{equation}
Therefore strong dephasing does not simply add noise after squeezing has formed but 
it suppresses the formation of squeezing itself.

The negative initial slope, as given in Eq.\eqref{slopedephasing}, establishes that dephasing \textit{does not prevent} the onset of Kerr squeezing. However, as also discussed in the previous section, it does not determine whether appreciable squeezing develops on the timescale predicted by the unitary evolution. To address this, we follow the same approach as for the zero-temperature damping case and retain the expression for the full dependence on  $R\equiv2N|\chi|t$.
\subsubsection{Finite-$R$ approximation for the dephasing channel}
We introduce again the variable $R\equiv N\tau= 2N|\chi|t$ and $ q\equiv \frac{\gamma_{\mathrm{deph}}}{2|\chi|}$. Then, in the large-$N$ limit at fixed \textit{finite} $R$ and $q$, the minimal variance \eqref{minimalvARIANCEdefinition}, for the dephasing case, has the expansion
\begin{equation}
V_-(t)=F_q(R)+\frac{1}{N}G_q(R)+\mathcal{O}(N^{-2})
\end{equation}
where the functions $F_q(R)$ and $G_q(R)$ are given by
\begin{align}
 F_q(R)&=1+2R(R+q)-2R\sqrt{1+(R+q)^2} \\   G_q(R)&=-R^2(R+q)^2+\frac{R^2\left[3(R+q)^3+5R+3q\right]}{\sqrt{1+(R+q)^2}}
\end{align}
It is straightforward to check that for $\gamma_{\mathrm{deph}}=0 \Leftrightarrow q=0$, we recover our previous unitary result \cite{KoppFragkosPikovski2022}. Now, we would like to ask how the dephasing, as parametrized by a $q\neq 0$, affects the variance in the finite-$R$ regime. Differentiating the function $F_q(R)$, we get
\begin{equation}
\frac{\partial}{\partial q}F_q\bigg|_{R=\rm const}=2R\biggl[1-\frac{R+q}{\sqrt{1+(R+q)^2}}\biggr]>0
\end{equation}
Therefore, at the same $R$, more dephasing means a larger variance. In other words, dephasing weakens the squeezing relative to the unitary prediction. However, a larger variance does not necessarily mean that squeezing has disappeared. This happens only once the variance reaches or exceeds 1, i.e $V_-(t)\geq 1$.
 To see this explicitly, we can equivalently rewrite $F_q(R)$ as
\begin{equation}
F_q(R)=1-\frac{2R}{\sqrt{1+(R+q)^2}+R+q}
\end{equation}
and for $R>0$ and finite $q\geq 0$, we get that $F_q(R)<1 \leftrightarrow V_-(t)<1$. Given also that the condition $\frac{G_q(R)}{N}\ll 1-F_q(R)$ is satisfied, we can conclude that squeezing is still present.

Now, let's assume that the coupling to the environment is weak, namely $q\ll 1$. In this case, Taylor expanding $F_q(R)$ around $q=0$ gives $F_q(R)=F_0(R)+q\frac{\partial F_q}{\partial q}|_{q=0}+\mathcal{O}(q^2)$. Using $q\equiv \frac{\gamma_{\rm deph}}{2|\chi|}$, we get
\begin{equation}\label{centraleq.deph}
F_{q\neq0}(R)-F_{q=0}(R)=    \frac{\gamma_{\rm deph}}{|\chi|}R\biggl(1-\frac{R}{\sqrt{1+R^2}}\biggr)+\mathcal{O}(q^2)
\end{equation}
This equation is central to our conclusion regarding the effect of dephasing on squeezing: It shows the environmental correction at times when unitary variance  is already substantially below vacuum noise, namely $V_-^{q=0}(t)<1$. From equation \eqref{centraleq.deph}, at $R=\mathcal{O}(1)$ the function $R\bigl(1-\frac{R}{\sqrt{1+R^2}}\bigr)$ is a finite numerical factor and therefore, the correction due to dephasing is small when the condition
\begin{equation}
\gamma_{\rm deph}\ll |\chi|
\end{equation}
is met. Therefore, this discussion establishes that at the time of appreciable squeezing, that is when $R=\mathcal{O}(1)$, when $N\gg1$ and $\gamma_{\rm deph}\ll |\chi|$ the axion mode develops squeezing close to the unitary prediction \cite{KoppFragkosPikovski2022}.\footnote{Recall that eq. \eqref{minimalVDephasing} is valid in the $R\ll1 $ limit and as such it is not the appropriate approximation to study squeezing at longer times.}

The inherent gravitational self-interaction of axions continuously tries to shear the initial coherent Hartree mode into a squeezed state \cite{KoppFragkosPikovski2022}. On the other hand, dephasing continuously randomizes the phase of the number-state components. Although dephasing modifies the initial squeezing rate already at linear order in time (see eq. \eqref{eqVminusDephasing}), it does not eliminate the onset of squeezing for any finite $\gamma_{\mathrm{deph}}>0$. The finite-R analysis in this section, establishes that, when $\gamma_{\mathrm{deph}}\ll|\chi|$, appreciable squeezing develops close to the unitary prediction on the timescale $R=2N|\chi|t=O(1)$. When $\gamma_{\mathrm{deph}}\sim|\chi|$, the variance at a given $R$ is modified and reaching a specified squeezing level can be delayed. In the strong-dephasing regime, that is the case when $\gamma_{\mathrm{deph}}\gg|\chi|$, the magnitude of the initial slope is reduced to approximately $4N\chi^2/\gamma_{\mathrm{deph}}$; Therefore, appreciable squeezing on the unitary formation timescale is suppressed, although a small initial reduction below vacuum remains for every finite dephasing rate.

In the previous paragraphs, we discussed squeezing during the onset ($R\ll 1$) as well as at timescales $R=\mathcal{O}(1)$. At much later times, $R\gg1$, dephasing also affects the maximum squeezing that can be reached. Preserving the unitary maximum, (see eqs. 6 and 7 in \cite{KoppFragkosPikovski2022}), requires $q\ll N^{-1/6}$ or equivalently $\gamma_{\mathrm{deph}}\ll 2|\chi|N^{-1/6}$. This condition is more stringent than $\gamma_{\rm deph}\ll |\chi|$. Consequently, dephasing that barely affects squeezing during its onset and the initial stages $R=\mathcal{O}(1)$ can still substantially reduce its maximal depth. In the dephasing-dominated regime, $N^{-1/6}\ll q\ll N^{1/4}$, the location and value of the minimum variance are approximately
\begin{equation}\label{eq:dephasingMaxima}
 R_{\max}^{(\mathrm{deph})}\simeq
\left(\frac{qN}{8}\right)^{1/5},
\qquad
V_{\min}^{(\mathrm{deph})}\simeq
\frac{5}{2^{7/5}}q^{4/5}N^{-1/5}.   
\end{equation}
Thus, increasing dephasing raises the minimum variance and shifts it to later times. Nevertheless, the minimum remains well below vacuum noise in this regime, allowing substantial squeezing to survive even when its maximal depth is determined by dephasing. The derivation and validity conditions are presented in Appendix \ref{app:dephasing}.

\subsection{Coupling of Axions to a finite-temperature Environment}
We now consider the Kerr unitary dynamics coupled to a finite-temperature environment. The Lindblad equation in this case is
\begin{equation}\label{eq:nonzerotempmaster}
\frac{d}{dt}\hat{\rho}=-i[\hat{H},\hat{\rho}]+\gamma\bigl(\bar n_{\rm th}+1\bigr)\mathcal{D}[\hat a]\hat \rho+\gamma\bar n_{\rm th}\mathcal{D}[\hat a^{\dagger}]\hat \rho
\end{equation}
where the dissipator is defined previously in \ref{sec:bridgingtwomodeLindblad}. The term $\propto \mathcal{D}[\hat a]\hat \rho$ describes loss of axions to the thermal bath, whereas the term $\propto \mathcal{D}[\hat a^{\dagger}]\hat \rho$ describes thermal excitation of the axion mode by the thermal environment.
The general solution, for an arbitrary initial state, was obtained in \cite{ChaturvediSrinivasan1991,Chaturvedi1999QuantumPS}. 

To compute the minimal variance $V_-(t)$, we introduce the Husimi Q-function, defined as $Q(\alpha,\alpha^*,t)=\langle\alpha|\hat\rho(t)|\alpha\rangle$, where $|\alpha\rangle$ is a coherent state \cite{Husimi1940}. This nonnegative function, normalized as $\frac{1}{\pi}\int d^2\alpha Q(\alpha,\alpha^*)=1$, provides a one-to-one representation of the density operator in phase space and belongs to the family of quantum quasi-probability distributions \cite{Kano1965,CahillGlauber1969,VannerPikovskiKim2015,GoldbergKlimovGrassl2020}. Antinormally ordered moments are obtained directly from $Q$ and the moments needed for $V_-(t)$ follow using the bosonic commutation relation. This provides a convenient route to deriving the short-time expansion of the minimal variance.  
\subsubsection{Husimi Q-function and Minimal variance for the onset of squeezing}
We consider a Kerr oscillator (nonlinearity $\chi$) initially in a coherent state $\ket{\alpha_0}$, coupled to a thermal bath of harmonic oscillators with mean occupation $\bar n_{\mathrm{th}}$. The Husimi $Q$-function is given by \cite{ChaturvediSrinivasan1991,Chaturvedi1999QuantumPS,Fragkos2025Thesis}
\begin{align}\label{eq:HusimiQ}\nonumber
Q(\alpha,\alpha^*,t)=\sum_{m,n=0}^{\infty} \frac{\bigl(\alpha^*\alpha_0 e^{-i(\omega-2\chi)t}\bigr)^m}{{m!}}\frac{\bigl(\alpha\alpha_0^*e^{+i(\omega-2\chi)t}\bigr)^{n}}{n!}e^{\frac{1}{2}\gamma t}\biggl[\sqrt{\Gamma_3(m-n)}\biggr]^{(m+n+1)}\times&\\ \times  \exp\biggl[\Bigl(\Gamma_{-}(m-n)-1\Bigr)|\alpha_0|^2\biggr]\cdot\exp\biggl[\Bigl(\Gamma_{+}(m-n)-1\Bigr)|\alpha|^2\biggr]
\end{align}
For notational simplicity, we set $k\equiv m-n$ with the functions in \eqref{eq:HusimiQ} taking the form \cite{ChaturvediSrinivasan1991}
\begin{equation}\label{eq:Gammapm(k3e}
 \Gamma_\pm(k) = \frac{\gamma_\pm}{\phi(k)}\frac{1}{\bigl[\coth\phi(k)+C(k)\bigr]}  , \qquad  \sqrt{\Gamma_3(k)} = \biggl(\cosh\phi(k) + C(k)\sinh\phi(k)\biggr)^{-1}
\end{equation}
\begin{equation}\label{eq:gamma_3(k)}
\gamma_+=\gamma\bar n_{\mathrm{th}} t,\quad\;\gamma_-=\gamma(\bar n_{\mathrm{th}}+1)t, \quad  \gamma_3(k) = -\gamma t(2\bar n_{\mathrm{th}}+1) - 2i\chi k t 
\end{equation}
\begin{equation}\label{eq:c(k)andphi(k)}
C(k) = -\frac{\gamma_3(k)}{2\phi(k)}, \qquad\phi(k) = \frac{t}{2}\sqrt{\bigl(\gamma(2\bar n_{\mathrm{th}}+1)+2i\chi k\bigr)^2 - 4\gamma^2\bar n_{\mathrm{th}}(\bar n_{\mathrm{th}}+1)}   
\end{equation}
For $t=0$, we obtain the usual expression for the Husimi Q-function corresponding to an initial coherent state, that is $Q=e^{-|\alpha-\alpha_0|^2}$ \cite{WallsMilburn2008}.

Our goal is to study the effect of a thermal bath on the unitary Kerr squeezing. To this end, we compute again the minimal quadrature variance 
\begin{equation}\label{eq:definitionMinimalVariance}
V_{-}(t)=1+2\bigl(\langle a^\dagger a\rangle-\langle a^\dagger\rangle\langle a\rangle\bigr)-2\sqrt{\bigl(\langle a^2\rangle-\langle a\rangle^2\bigr)\bigl(\langle a^{\dagger2}\rangle-\langle a^\dagger\rangle^2\bigr)}   
\end{equation}
at early times, leading to an expression for $ V_{-}(t)$ up to, and including, order $t^2$. We first expand $Q(\alpha,\alpha^*,t)$ in powers of $t$. In doing so, see details Appendix \ref{app:Qfunction}, we end up with the expression\footnote{Notice that in this subsection we denote $N_0$ the initial mean occupation of the selected axion mode.}
\begin{equation}\label{vminustemp}
\boxed{V_-(t)=1+\left(2\gamma\bar n_{\mathrm{th}} -4|\chi|N_0\right)t+
\left[8\chi^2N_0^2-\gamma^2\bar n_{\mathrm{th}}+4\gamma|\chi|N_0(1-\bar n_{\mathrm{th}})\right]t^2
+O(t^3)}
\end{equation}
This result, eq.\eqref{vminustemp}, reduces to previously considered cases:\\
$\bullet$ For $\bar n_{\mathrm{th}}=0$,
\begin{equation}
 V_-^{\bar n_{\mathrm{th}}=0}(t)=1-4|\chi|N_0t+
\left[8\chi^2N_0^2+4\gamma|\chi|N_0\right]t^2+O(t^3)   
\end{equation}
which agrees with relation \eqref{Vminusdissip} previously obtained.\\\\
$\bullet$ For $\chi=0$, eq. \eqref{vminustemp} becomes
\begin{equation}
V^{\chi=0}_-(t)=1+2\gamma\bar n_{\mathrm{th}}t-\gamma^2\bar n_{\mathrm{th}}t^2+O(t^3)  
\end{equation}
which matches exactly the short-time expansion of the expected result for a linear damped oscillator coupled to a thermal bath, that is $V_-(t)=V_+(t)=1+2\bar n_{\mathrm{th}}\left(1-e^{-\gamma t}\right)$ \cite{WallsMilburn2008,WallsMilburn1985DissipationCoherence,MilburnWalls1983DampedOscillator,Phoenix1990DampedOscillator}.\\\\
$\bullet$ For $\chi=\bar n_{\mathrm{th}}=0$ we get $V_-(t)=1$ which is compatible with the fact that in absence of self-Kerr and temperature, the dissipation does not produce any squeezing and the initial coherent state, in a rotating frame, evolves as $|\alpha\rangle_t=|\alpha_0e^{-\gamma t/2}\rangle$ \cite{WallsMilburn2008}.
\subsubsection*{\underline{Physical content of the result \eqref{vminustemp}}}
Before discussing the physics described in \eqref{vminustemp}, let us first briefly recap the following key point from \cite{KoppFragkosPikovski2022}: there, Kerr squeezing is generated by the intrinsic nonlinearity of the self-gravitating axion field. Terms proportional to $|\chi|$ in \eqref{vminustemp} originate precisely from this nonlinear dynamics. As we have seen earlier, damping of axions into a zero-temperature bath of SHOs \textit{does not} affect the onset of squeezing at order $\mathcal{O}(t)$. Effects of damping into a zero-temperature environment appear only as a higher-order effect, at order $t^2$. As a result, initial squeezing found in \cite{KoppFragkosPikovski2022} is robust with respect to this dissipative channel.

In contrast, and similarly to the dephasing channel, equation \eqref{vminustemp} reveals that the effect of temperature already enters at the linear level, that is at order $\mathcal{O}(t)$. This means that as soon as the squeezing due to nonlinearities kicks in, immediately there is a competition with the thermal environment. This competition is evident from the opposite signs of the two contributions: $+2\gamma \bar n_{\mathrm{th}}t$ versus $-4|\chi|N_0t$. At short times, Kerr evolution shears the initially circular quadrature fluctuations in phase space, reducing the variance along one direction below the vacuum limit, whereas thermal noise broadens the fluctuations and opposes this reduction. This can be immediately seen, at early times, from equation \eqref{vminustemp}
\begin{equation}\label{eq:conditionOnsetTneq0}
    \frac{d}{dt}V_-(t)\biggl|_{t=0^+}=2\gamma\bar{n}_{\rm th}-4|\chi|N_0
\end{equation}
For $\gamma\bar{n}_{\rm th}<2|\chi|N_0$ the variance initially decreases below one which means that squeezing begins. When $\gamma\bar{n}_{\rm th}=2|\chi|N_0$ holds exactly, then the linear in time coefficient in eq. \eqref{vminustemp} vanishes, resulting in $V_-(t)=1+2\gamma|\chi|N_0t^2+\mathcal{O}(t^3)$. Thus, for positive values of $\gamma$, the variance initially rises above the vacuum value $V_-(t)=1$. Moreover, for $ \gamma\bar{n}_{\rm th}>2|\chi|N_0$ the variance initially increases indicating absence of squeezing sufficiently close to $t=0$. Finally, for $\gamma\bar{n}_{\rm th}\ll2|\chi|N_0$, squeezing begins at a rate close to that of the unitary evolution \cite{KoppFragkosPikovski2022}.

However, as discussed in sections (\ref{sec:zeroTemp}) and \eqref{sec:Dephasing}, even a negative initial slope does not establish that the variance will fall substantially below one. For this, we need to use again the finite-R calculation.

\subsubsection{Finite-R approximation for the thermal environment}
We introduce once again the same scaling variables as in the unitary Kerr problem, $\tau\equiv2|\chi|t$ and $R \equiv N_0\tau=2N_0|\chi|t$, while taking the limits $N_0\rightarrow \infty,\quad\tau\ll 1,\quad R=\text{finite}$. This way, we retain the full dependence on $R$ and do not require $R\ll 1$. This allows us to follow the development of significant squeezing at $R=\mathcal{O}(1)$. For weak thermal damping, that is assuming $\gamma \ll N_0|\chi|$ and $\gamma\bar{n}_{\rm th}\ll N_0|\chi|$, the finite-$R$ result can be written perturbatively as
\begin{equation}
\boxed{
\label{thermal_finite_R_full}
\begin{split}
V_-(R)&=\underbrace{V_R^{(0)}(\gamma=0)}_{\rm Unitary} +  \underbrace{V_R(\gamma\neq 0)}_{(T=0)-\rm Losses}+\underbrace{V_{R}^{\rm th}(\bar{n}_{\rm th}\neq 0)}_{(T\neq0)-\mathrm{Contribution}}+ \mathcal{O}\biggl[\frac{1}{N_0^2},\frac{1}{N_0}\frac{\gamma(1+\bar{n}_{\rm th})}{N_0|\chi|},\frac{\gamma^2(1+\bar{n}_{\rm th})^2}{N_0^2|\chi|^2}\biggr]
\\[4pt]
\end{split}}
\end{equation}
where 
\begin{equation}
V_R^{(0)}(\gamma=0)=\big(\sqrt{1+R^2}-R\bigr)^2   +\frac{1}{N_0}\biggl(\frac{R^3(5+3R^2)}{\sqrt{1+R^2}}
-R^4\biggr) 
\end{equation}
is the unitary, finite-R result \cite{KoppFragkosPikovski2022}, the function $V_R(\gamma\neq 0)$ 
\begin{equation}
V_R(\gamma\neq 0) = \frac{\gamma}{N_0|\chi|}\frac{1}{3}\left(\frac{3R^2+5R^4}{\sqrt{1+R^2}}-5R^3\right)
\end{equation}
is the temperature-independent piece matching exactly eq. \eqref{eq:T=0dissipationcorrection} and 
\begin{equation}
V_{R}^{\rm th}(\bar{n}_{\rm th}\neq 0)=\frac{\gamma \bar n_{\mathrm{th}} }{N_0|\chi|}R\,
\biggl(1+\frac{2R^2}{3}\biggr)\biggl(1-\frac{R}{\sqrt{1+R^2}}\biggr)
\end{equation}
is the additional thermal contribution.
\subsubsection*{\underline{Physical Interpretation of Eq.\eqref{thermal_finite_R_full}}:}
\begin{figure}[t]
\centering
\includegraphics[
width=0.9\linewidth,
height=0.35\textheight,keepaspectratio
]{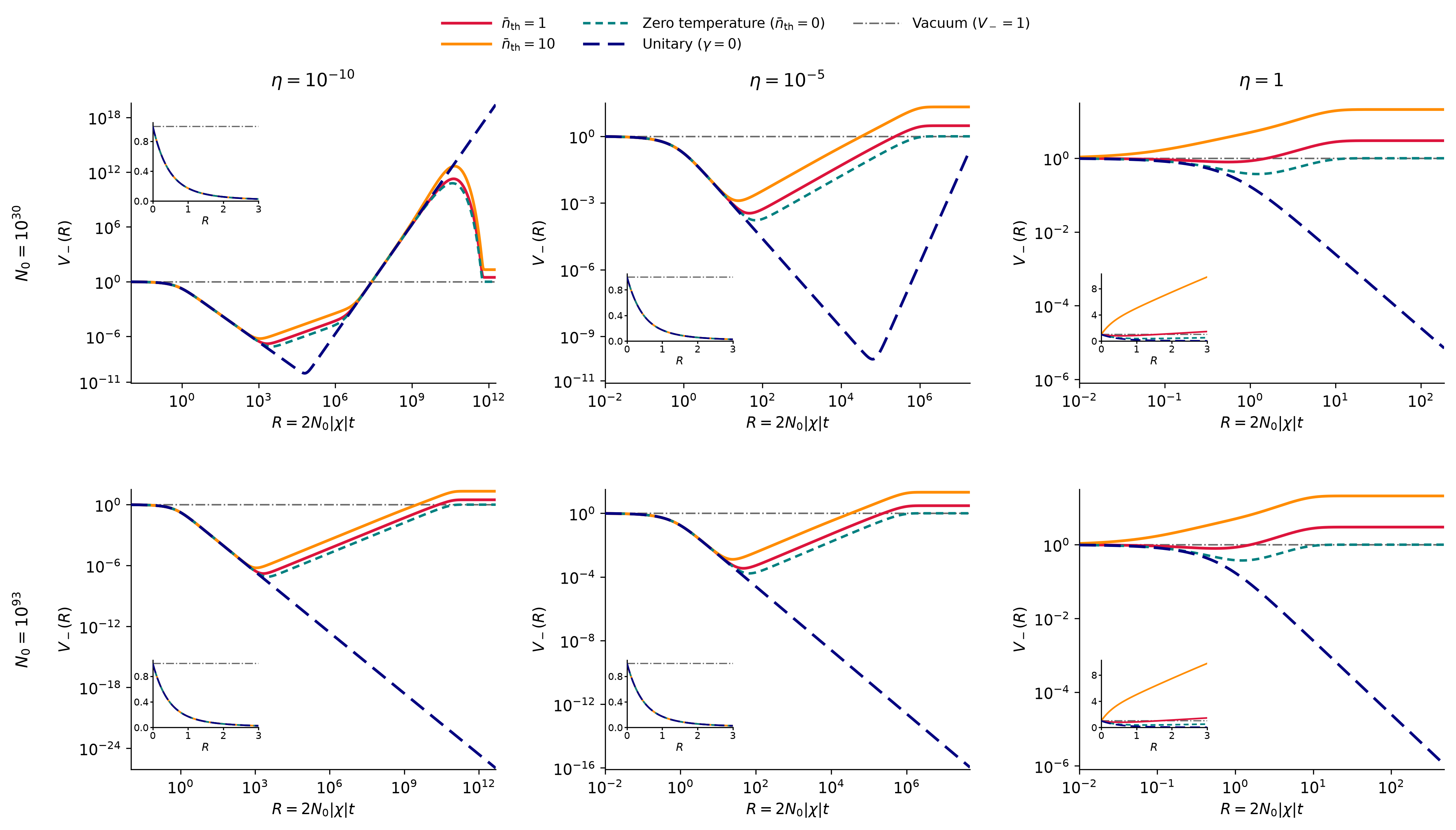}
\caption{\small Minimum quadrature variance $V_-(t)$ at all times, for an initially coherent mode coupled to a finite-temperature environment. Here, $\eta=\frac{\gamma}{N_0|\chi|}$ with $N_0$ the initial mean number of axions in the mode. Weak damping and thermal noise preserve the near-unitary onset while increasing thermal occupation  makes squeezing shallower and shorter-lived. The initial decrease below the vacuum level requires $\eta\bar n_{\rm th}<2$.}
\label{fig:squeezingcosmology}
\end{figure}
The environmental correction contains two distinct contributions: the zero-temperature loss term, controlled by $\gamma/(N_0|\chi|)$, and the additional thermal-noise term, controlled by $\gamma\bar n_{\mathrm{th}}/(N_0|\chi|)$. Both contributions are positive for $(R>0)$, which is always the case, and therefore increase the minimal variance relative to the unitary prediction within the present approximation. For $R\sim1$, sufficiently large $N_0$, $\gamma/(N_0|\chi|)\ll1$ and $\gamma\bar n_{\mathrm{th}}/(N_0|\chi|)\ll1$, these corrections remain small and appreciable squeezing develops close to the unitary prediction. Therefore, the conditions that establish a sufficient regime for squeezing to remain close to the unitary prediction are
\begin{equation}
\gamma/(N_0|\chi|)\ll1,  \qquad  \gamma\bar n_{\mathrm{th}}/(N_0|\chi|)\ll1
\end{equation}
Physically, these conditions ensure that neither depletion nor accumulated thermal noise substantially modifies the formation of squeezing. This finite-time robustness statement is stronger than the condition for immediate onset \eqref{eq:conditionOnsetTneq0} obtained from the initial slope. When $\gamma\bar n_{\mathrm{th}}/(N_0|\chi|)\sim 1$, thermal noise can substantially modify the squeezing dynamics, and the evolution requires a treatment beyond the present first-order environmental expansion.

\subsection{Astrophysical application: Environmental loss in an axion solitonic core}
A concrete application of our results in this section is provided by an ultralight axion solitonic core.\footnote{For the astrophysical estimates here, we restore the units of $\hbar$ and $c$.} Recall that our core result in \cite{KoppFragkosPikovski2022}, is that the inherent gravitational interactions generate quadrature squeezing, below the vacuum limit, when an initially coherent axion mode within the single-mode Hartree approximation is considered. For a solitonic core hosted by a halo of mass $M_h=2\times10^{12}M_\odot$, the squeezing time, normalized with respect to the ultralight mass $m=\mathcal{O}(10^{-22}\rm eV/c^2)$, is\footnote{See Table I and Appendix H in Ref.\cite{KoppFragkosPikovski2022}.} 
\begin{equation}\label{eq:Solitonsqztimescale}
t_{\mathrm{sqz}}\simeq4.4\cdot 10^6 \biggl(\frac{mc^2}{10^{-22}\rm eV}\biggr)^{-1}\mathrm{yr},
\qquad
 N|\chi|
=\frac{\sinh1}{2t_{\mathrm{sqz}}}
\simeq4.2\times10^{-15}\biggl(\frac{mc^2}{10^{-22}\rm eV}\biggr)\mathrm{s}^{-1},
\end{equation}
where the mass of the axion is a free parameter. Here, $t_{\rm sqz}$ is the unitary squeezing reference time defined as $R(t_{\rm sqz})=\sinh 1$, as in Ref. \cite{KoppFragkosPikovski2022}. The solitonic core estimate in Eq.\eqref{eq:Solitonsqztimescale} is within the ultralight axion regime. For comparison, the cosmological Hubble-volume estimate of Ref. \cite{KoppFragkosPikovski2022} gives $t_{\rm sqz}\simeq 66\rm \mu s$ for $m\simeq 10^{-5} \rm eV/c^2$. 

In this work, our analysis extends the unitary result by allowing the selected Hartree mode to interact with an environment and thus lose coherence. For example, considering a zero-temperature environment, in the weak loss and large $N$-limit, the finite-time variance correction due to the environmental coupling, as given by eq. \eqref{eq:T=0dissipationcorrection}, is
\begin{equation}
\Delta_{\gamma}V_-(t)\simeq\frac{\gamma}{N|\chi|}\frac{1}{3}\left(\frac{3R^2+5R^4}{\sqrt{1+R^2}}-5R^3\right)
\end{equation}
This expression, when evaluated at the onset of squeezing \eqref{eq:Solitonsqztimescale}, leads to 
\begin{equation}
\Delta_{\gamma}V_-\left(t_{\mathrm{sqz}}\right)
\simeq\frac{\gamma}{4N|\chi|}.    
\end{equation}
Thus the relevant robustness condition is, as discussed in Sec.\ref{sec:zeroTemp}, $\gamma\ll N|\chi|$. For example, requiring the environmental correction to remain below $\sim10\%$ of the unitary variance gives 
\begin{equation}
\gamma\lesssim2.3\times10^{-16}\biggl(\frac{mc^2}{10^{-22}\rm eV}\biggr)\mathrm{s}^{-1}    
\end{equation}
As an example of a possible astrophysical environment, Davies and Mocz \cite{Davies_2020} studied fuzzy-dark-matter solitons around a supermassive black hole. Their setup consists of a gravity dominated ultralight soliton, which in the non-relativistic limit is described by the Schrodinger-Poisson equation. Their calculation provides an estimate of axion absorption by the black hole, which can be mapped onto an \textit{effective one-way loss rate for the solitonic mode}. Using their soliton profile and the  Hui–Unruh absorption estimate \cite{HuiOstrikerTremaineWitten2017,Unruh1976Absorption}, one obtains,
\begin{equation}\label{BHabsorption}
\gamma_{\rm BH}\equiv -\frac{\dot M_s}{M_s}=
\frac{\dot M_\bullet}{M_s}
\simeq \frac{16\pi G^2M_\bullet^2\rho_c}{c^3M_s}F(\xi)
\end{equation}
Here, \(\dot M_\bullet\) denotes the contribution to black-hole growth from absorption of the soliton. Writing $M_s(t)=mN(t)$, eq.\eqref{BHabsorption}  implies $\dot N=-\gamma_{\rm BH}N$. We model this absorption as an effective Markovian one-way particle-loss channel. The absorption factor is $F(\xi)=\frac{\xi}{1-e^{-\xi}},\quad\xi=\frac{2\pi GM_\bullet m}{\hbar\upsilon}\simeq19.07\frac{M_\bullet}{M_s},$
where \(\upsilon\) is the soliton virial velocity.

We take $M_\bullet=4\cdot10^6 M_\odot $ and $M_s=1.6\cdot 10^{9}(mc^2/10^{-22}\rm eV)^{-1}M_\odot$ the black hole and solitonic core masses, respectively. The central core density scaling of the solitonic core is approximately $\rho_c\simeq50\cdot(mc^2/10^{-22}\rm eV)^{2}\cdot M_\odot\cdot\rm (pc)^{-3}$ \cite{Davies_2020}.  For the case of ultralight axions $mc^2=\mathcal{O}(10^{-22})\rm eV$, we have that $\xi\simeq 0.048\ll1$, and we get
\begin{equation}
\gamma_{\rm BH}\simeq6\times10^{-28}
 {\rm s}^{-1}
\end{equation}
and
\begin{equation}
    \frac{\gamma_{\rm BH}}{N|\chi|}\simeq 10^{-13}
\end{equation}
Therefore, the resulting correction to the variance is
\begin{equation}
\Delta_{\gamma}V_-\left(t_{\mathrm{sqz}}\right)\simeq 10^{-14} 
\end{equation}
Within the ultralight regime for which the soliton and black-hole model is applicable, this correction is far below the $10\%$ bound. Therefore, in this illustrative estimate, black-hole absorption, or equivalently losses of the axion solitonic core, has a negligible effect on the formation of squeezing found in \cite{KoppFragkosPikovski2022}. In other words, the presence of a black hole does not hinder the development of solitonic squeezing.

\section{Conclusions}
\label{sec:conclusions}
Our previous work showed that an initially coherent axion mode develops quadrature squeezing through its gravitational self-interaction, within the single-mode Hartree approximation \cite{KoppFragkosPikovski2022}. The inherent gravitational self-interaction that drives the classical mean-field evolution also squeezes the quantum fluctuations, reducing the variance of one quadrature below the vacuum level. Because squeezing originates from the gravitational self-interactions alone, this same mechanism can also operate in different light scalar dark matter models. 
In other words, this mechanism is not specific to the QCD-axion physics and is an omnipresent feature of any self-gravitating light scalar dark matter, resulting in quantum squeezed states of light scalar dark matter. Specifically, for the highly occupied axion fields considered, squeezing emerges on cosmologically short timescales and can subsequently reach very large values. For example, the cosmological estimate gives $t_{\rm sqz}\simeq 66\mu s$ for QCD-axion mass $mc^2\simeq 10^{-5\rm}\rm eV$.

Here, we examined whether environmental interactions can destroy this quantum feature, considering three representative Markovian channels: Zero-temperature loss, thermal dissipation and pure dephasing. Our results show that squeezing can remain robust against such decoherence. Zero-temperature loss leaves the initial squeezing rate unchanged. Thermal noise competes with squeezing from the outset and can prevent its initial onset, whereas pure dephasing reduces the initial squeezing rate without eliminating onset at any finite dephasing rate. Appreciable squeezing remains close to the unitary predictions in the regimes identified in Sec.\ref{sec:Decoherence of ADM} where the relevant environmental effects are sufficiently weak.

These findings describe a competition between environmental effects and squeezing generated dynamically by the field’s intrinsic gravitational self-interaction. Suppressing appreciable squeezing requires sufficiently strong loss, thermal noise or dephasing relative to the corresponding squeezing scales. Squeezing is more robust against decoherence than other non-classical states, such as spatial superpositions, and can survive environmental interactions over a finite period of the evolution. Our results support squeezing as a potentially robust companion of coherent light scalar dark matter, suggesting that quantum features might be persistent in such dark matter candidates.

\section{Acknowledgements}
This work was supported by the National Science Foundation under Grant No. 2239498, by NASA under Grant No. 80NSSC25K7051 and by the “Table-top experiments for fundamental physics” program sponsored by the Alfred P. Sloan Foundation, Gordon and Betty Moore Foundation, John Templeton Foundation, and Simons Foundation under Grant No. G-2023-21102.

\appendix
\label{sec:appendices}
\section{Time of maximal squeezing}
\subsection{Case I: Zero-temperature environment}\label{app:zero_Temp_tmax}
In unitary case, we have estimated the time of maximal squeezing, that is the time in which the minimal variance reaches its minimum value, is $N|\chi|t_{\mathrm{max}}\simeq0.3N^{\frac{1}{6}}$ (see eq. 7 in \cite{KoppFragkosPikovski2022}, or equivalently, applied to axions in a Hubble volume, given by eq.\eqref{maximumsqztime}). For $R\gg 1$ but finite, we get from equation \eqref{minimalvardissip}
\begin{equation}\label{zerotempvariancetmax}
V^{(\gamma)}_-(R)\simeq \underbrace{\frac{1}{4R^2}}_{\rm \text{Self-Kerr squeezing}}+\underbrace{\frac{2R^4}{N}}_{\rm \text{Self-Kerr correction}}+\underbrace{\frac{\gamma}{6N|\chi|} R}_{\rm Loss}  
\end{equation}
In the limit $\gamma\rightarrow0$, we recover the results in \cite{KoppFragkosPikovski2022,BajerMiranowiczTanas2002}.
The first term tends to dynamically decrease the minimal variance while the second and third terms increase with time\footnote{Recall that the parameter is proportional to time, i.e $R\propto t$.}. Physically, the first term is caused by the unitary self-Kerr non-linear evolution, the second term is a correction of the first term that becomes significant at later times, i.e larger values of $R$, while the third term accumulates vacuum noise due to the loss channel. Although the last two terms are initially small, they determine when the variance stops decreasing. Minimizing this expression we get 
\begin{equation}\label{eq:minimization}
\frac{dV^{(\gamma)}_{-}(R)}{dR}|_{R=R_{\rm max}^{(\gamma)}}=0\Longleftrightarrow 8\bigl(R_{\rm max}^{(\gamma)}\bigr)^6+\frac{\gamma}{6|\chi|}\bigl(R_{\rm max}^{(\gamma)}\bigr)^3-\frac{N}{2}=0
\end{equation}
Setting $y\simeq\bigl(R_{max}^{(\gamma)}\bigr)^3$ and solving the quadratic equation, we get
\begin{equation}\label{eq:RmaxZeroTemp}
R_{\rm max}^{(\gamma)}=\biggl[\frac{\sqrt{\bigl(\frac{\gamma}{6|\chi|}\bigr)^2+16N}-\frac{\gamma}{6|\chi|}}{16}\biggr]^{1/3} 
\end{equation}
Therefore,
\begin{equation}
t_{\mathrm{max}}^{(\gamma)}=\frac{R_{\rm max}^{(\gamma)}}{2N|\chi|}  
\end{equation}
For clarity, we stress here that this corresponds to an asymptotic value of the variance, at $R\gg1$, rather than the exact solution of the full dynamics. Nevertheless, this is adequate to estimate $t^{(\gamma)}_{\rm max}$. Inserting the value for $R_{\rm max}^{(\gamma)}$ back into \eqref{zerotempvariancetmax}, that is computing the minimum variance which occurs at $t_{\rm max}$, i.e  $V_{\rm min}\equiv V(R_{\rm max})$ we get
\begin{equation}
V_{\rm min}\equiv V(R_{\rm max})=\frac{6\bigl(R_{\rm max}^{(\gamma)}\bigr)^4}{N}+\frac{\gamma}{4N|\chi|}R_{\rm max}^{(\gamma)}
\end{equation} 
For $\gamma=0$, we recover the unitary results as obtained in \cite{KoppFragkosPikovski2022}
\begin{equation}
R^{(0)}_{\rm max}=2^{-2/3}N^{1/6}, \qquad t_{\mathrm{max}}^{(0)}\simeq \frac{2^{-5/3}N^{-5/6}}{|\chi|}, \qquad V^{(0)}_{\rm min}=\frac{3}{2^{5/3}}N^{-1/3} 
\end{equation}
Equation \eqref{eq:RmaxZeroTemp} reveals two regimes: For $\gamma\ll \sqrt{N}|\chi|$, $R_{\rm max}^{(\gamma)}$ becomes 
\begin{equation}
R_{\rm max}^{(\gamma)}\simeq 2^{-2/3}N^{1/6}\bigl(1-\frac{\gamma}{72|\chi|\sqrt{N}}\bigr)  
\end{equation}
and 
\begin{equation}
t_{\mathrm{max}}^{(\gamma)}\simeq t_{\mathrm{max}}^{(0)}  \biggl(1-\frac{\gamma}{72|\chi|\sqrt{N}}\biggr)
\end{equation}
Thus, the unitary values of the maximum are approximately preserved. For $|\chi|\sqrt{N}\ll \gamma \ll N|\chi| $, the maximum squeezing is determined by the loss. From \eqref{eq:minimization}, the loss term dominates, leading to 
\begin{equation}
R_{\rm max}^{(\gamma)}
\simeq\biggl[\frac{3N|\chi|}{\gamma}\biggr]^{1/3},  \qquad t_{\mathrm{max}}^{(\gamma)}\simeq\biggl(\frac{3}{8\gamma (N|\chi|)^2}\biggr)^{1/3}
\end{equation}
The minimum variance becomes 
\begin{equation}
V_{\rm min}\simeq \frac{3^{1/3}}{4}\biggl(\frac{\gamma}{N|\chi|}\biggr)^{2/3} 
\end{equation}
However, since we work in a regime where $\gamma\ll N|\chi|$, we conclude that the minimum variance is $ V_{\rm min}\simeq \frac{3^{1/3}}{4}\bigl(\frac{\gamma}{N|\chi|}\bigr)^{2/3} \ll 1$, namely substantial squeezing remains possible. However, it is weaker than the unitary case $V_{\rm min}\propto N^{-1/3}$ and it is also reached earlier. We refer the reader to Fig.\ref{fig:zeroTemp} in the main text.

As a final remark, in the left-panel of Fig.\ref{fig:zeroTemp}, the late time behavior is governed by the competition between the self-Kerr correction term $2R^4/N$ and loss-term $\gamma R/6N|\chi|$, as given by Eq.\eqref{zerotempvariancetmax}. Because $2R^4/N$ grows much faster in time, it eventually overtakes the loss contribution for an occupation of $N=10^{30}$, leading to a steep rise beyond the vacuum variance level, i.e $V\geq 1$. For the much larger occupation of $N=10^{93}$, the self-Kerr correction term is strongly suppressed, and damping reduces the nonlinear evolution before this distortion becomes appreciable. This is the reason why in the large-R limit, for $N=10^{93}$, the exact variance \eqref{eq:MinimalvarianceT=0atalltimes} approaches one from below. 

\subsection{Case II: Dephasing environment}\label{app:dephasing}
To obtain $t_{\max}$ for the dephasing environment, we follow the same logic as in the unitary and zero-temperature environment case. We introduce the variables $R\equiv N\tau= 2N|\chi|t$ and $q\equiv \frac{\gamma_{\mathrm{deph}}}{2|\chi|}$. In the large-$N$ limit at fixed finite $R$ and fixed $q$, the minimal variance, as also discussed in \ref{sec:Dephasing} for the dephasing case, has the expansion
\begin{equation}
V_-(t)=F_q(R)+\frac{1}{N}G_q(R)+O(N^{-2})
\end{equation}
where
\begin{equation}
F_q(R)=1+2R(R+q)-2R\sqrt{1+(R+q)^2}
\end{equation}
and
\begin{equation}
G_q(R)=-R^2(R+q)^2+\frac{R^2\left[3(R+q)^3+5R+3q\right]}{\sqrt{1+(R+q)^2}}
\end{equation}
It is straightforward to check that for $\gamma_{\mathrm{deph}}=0 \Leftrightarrow q=0$, we recover the unitary result \cite{KoppFragkosPikovski2022}. Taking the derivative 
\begin{equation}
\frac{\partial F_{q}(R)}{\partial R}\bigg|_{R=0}^{q=\rm fixed}=-2\bigl(\sqrt{1+q^2}-q\bigr)<0
\end{equation}
This analytical result is depicted by the insets in Fig.\ref{fig:DephasingLindblad}. This implies that squeezing starts at every $q>0$, but increasing $q$ reduces the magnitude of the initial decrease. Moreover, $F_q(R)$ does not contain $N$. Therefore, the initial evolution is essentially identical in the two rows at the same q.

For $R\gg 1$, $R\gg q$ and $\frac{R^2}{N}\ll1$, we get
\begin{equation}
V_-^{(\mathrm{deph})}(R)
\simeq\underbrace{\frac{1}{4R^2}}_{\text{unitary squeezing}}+
\underbrace{\frac{q}{R}}_{\text{dephasing contribution}}
+\underbrace{\frac{2R^4}{N}}_{\text{finite-}N\text{correction}}
\end{equation}
The first two terms decrease with time whereas the last one increases. Their competition determines the maximum squeezing that can be reached. Differentiating with respect to $R$, we get
\begin{equation}\label{eq:minimazedephasing}
\left.\frac{d}{dR} V_-^{(\mathrm{deph})}(R)
\right|_{R=R_{\mathrm{max}}^{(\gamma_{\mathrm{deph}})}}=0 \Longrightarrow \frac{8}{N}\bigl(R^{(\gamma_{\mathrm{deph}})}_{\max}\bigr)^6-qR^{(\gamma_{\mathrm{deph}})}_{\max}-\frac{1}{2}\simeq 0
\end{equation}
Before discussing the maximum, this relation explains analytically the behavior in Fig.\ref{fig:DephasingLindblad}. Taking the derivative,  
\begin{equation}
\frac{dR^{(\gamma_{\mathrm{deph}})}_{\max}}{dq}\bigg|_{N- \rm fixed}=\frac{\bigl(R^{(\gamma_{\mathrm{deph}})}_{\max}\bigr)^2}{5qR^{(\gamma_{\mathrm{deph}})}_{\max}+3}>0
\end{equation}
implies increasing dephasing $q$ moves the minimum variance to larger $R$. This behavior can also be seen in Fig.\ref{fig:DephasingLindblad} (orange curves at the rightmost part). Moreover,
\begin{equation}
\frac{d}{dq}V_{\rm min}=\frac{\partial}{\partial q}V_{-}\bigg|_{R^{(\gamma_{\mathrm{deph}})}_{\max}}=\frac{1}{R^{(\gamma_{\mathrm{deph}})}_{\max}}>0    
\end{equation}
shows that the minimum becomes higher.

The time of maximal squeezing can be found by minimizing $V_-^{(\mathrm{deph})}(R)$ with respect to $R$. Solving \eqref{eq:minimazedephasing}  for $R_{\mathrm{max}}^{(\gamma_{\mathrm{deph}})}$ and inserting back into $V_-^{(\mathrm{deph})}(R)$, we get 
\begin{equation}\label{eq:VminDephAppendix}
V_{\rm min}^{(\rm deph)}\simeq \frac{3}{8\bigl(R_{\mathrm{max}}^{(\gamma_{\mathrm{deph}})}\bigr)^2}+\frac{5q}{4R_{\mathrm{max}}^{(\gamma_{\mathrm{deph}})}}
\end{equation}
Equation \eqref{eq:minimazedephasing} determines $R_{\mathrm{max}}^{(\gamma_{\mathrm{deph}})}$ and hence $t_{\mathrm{max}}^{\mathrm{deph}}$. Equation \eqref{eq:VminDephAppendix} gives the corresponding minimum variance. In particular,
Eq.\eqref{eq:VminDephAppendix} gives two distinct regimes: 1) $q\ll N^{-1/6}$ \quad and \quad 2) $N^{-1/6}\ll q \ll N^{1/4}$.\\\\$\bullet$ For $q\ll N^{-1/6}$, or equivalently $\gamma_{\mathrm{deph}}\ll2|\chi|N^{-1/6}$, the term $qR^{(\gamma_{\mathrm{deph}})}_{\max}$ in Eq.\eqref{eq:minimazedephasing} is negligible and therefore, it becomes $\frac{8}{N}\bigl(R^{(\gamma_{\mathrm{deph}}\rightarrow 0)}_{\max}\bigr)^6\approx \frac{1}{2}$. Thus, we recover the unitary results $ R^{(0)}_{\max}\simeq2^{-2/3}N^{1/6}$, $V_{\rm min}^{(0)}=\frac{3}{2^{5/3}}N^{-1/3}$
and $t_{\max}^{(0)}\simeq2^{-5/3}\frac{N^{1/6}}{N|\chi|} \Longleftrightarrow N|\chi|t_{\max}^{(0)}\simeq2^{-5/3}N^{1/6}\simeq0.3\,N^{1/6}$ \cite{KoppFragkosPikovski2022}.\\\\$\bullet$ For $N^{-1/6}\ll q \ll N^{1/4}$, the scaling changes. Equation \eqref{eq:minimazedephasing} now is approximately equal to
\begin{equation}
\frac{8}{N}\bigl(R^{(\gamma_{\mathrm{deph}})}_{\max}\bigr)^6\approx qR^{(\gamma_{\mathrm{deph}})}_{\max}
\end{equation}
leading to $R_{\max}^{(\mathrm{deph})}\sim\left(\frac{qN}{8}\right)^{1/5}$. The time at which squeezing is maximal is directly given from the definition of $R$ to give $t_{\max}^{(\mathrm{deph})}\equiv\frac{R^{(\gamma_{\mathrm{deph}})}_{\max}}{2N|\chi|}$ and therefore $t_{\max}^{(\mathrm{deph})}\sim\frac{1}{2N|\chi|}\left(\frac{\gamma_{\mathrm{deph}}}{|\chi|}\frac{N}{16}\right)^{1/5}$. The minimum value that the variance reaches in that case is 
\begin{equation}
V_{\min}^{(\mathrm{deph})}\simeq\frac{5}{2^{11/5}}
\left(\frac{\gamma_{\mathrm{deph}}}{|\chi|}\right)^{4/5}N^{-1/5}
\end{equation}
Therefore, in this case, dephasing determines the minimal variance. Nevertheless, it can remain well below the vacuum level, $V_{\min}^{(\mathrm{deph})}\ll 1$, allowing substantial squeezing to survive.

\section{Short-time expansion of the thermal $Q$-function}
\label{app:Qfunction}

First, in order to make the notation more compact, we define $x\equiv \alpha^*\alpha_0$, $y\equiv \alpha\alpha_0^*$, $N_0\equiv |\alpha_0|^2$ and $m-n=k$. 

We also define $A(k)=\gamma(2\bar n_{\mathrm{th}}+1)+2i\chi k$ and $\Delta(k)=A(k)^2-4\gamma^2\bar n_{\mathrm{th}}(\bar n_{\mathrm{th}}+1)$. Both $A(k)\equiv A(k,\chi,\gamma,\bar n_{\mathrm{th}})$ and $\Delta(k)\equiv\Delta(k,\chi,\gamma,\bar n_{\mathrm{th}})$ are \textit{independent of time} $t$ and depend only on the physical parameters of the problem. 

Then, from \eqref{eq:c(k)andphi(k)}, we have $\phi(k)=\frac{t}{2}\sqrt{\Delta(k)}\equiv \phi_0(k)\,t$, where $\phi_0(k)=\frac12\sqrt{\Delta(k)}$ is time-independent. Moreover, from \eqref{eq:gamma_3(k)}, $\gamma_3(k)=-A(k)t$. Therefore, equation \eqref{eq:c(k)andphi(k)}: $C(k)=A(k)/(2\phi_0(k))$ is \textit{independent} of $t$.

Expanding the $\Gamma$-functions for small $t$, we get
\begin{equation}\label{eq:GammaFunctionsTaylor}
\begin{aligned}
\Gamma_3(k) &=1-A(k)t+\left[\frac{A(k)^2}{2}+\gamma^2\bar n_{\mathrm{th}}(\bar n_{\mathrm{th}}+1)\right]t^2+O(t^3)
\\[4pt]
\Gamma_+(k) &= \gamma\bar n_{\mathrm{th}}\,t-\frac{\gamma\bar n_{\mathrm{th}}}{2}A(k)t^2+O(t^3)\\[4pt]
\Gamma_-(k) &=\gamma(\bar n_{\mathrm{th}}+1)t-\frac{\gamma(\bar n_{\mathrm{th}}+1)}{2}A(k)t^2+O(t^3) 
\end{aligned}
\end{equation}
Using \eqref{eq:GammaFunctionsTaylor}, we rewrite \eqref{eq:HusimiQ}
\begin{equation}
Q(\alpha,\alpha^*,t)=e^{-|\alpha-\alpha_0|^2}\left[1+t\,Q_1(\alpha,\alpha^*)+t^2\,Q_2(\alpha,\alpha^*)\right]+O(t^3)
\end{equation}
with
\begin{equation}
\begin{aligned}
Q_1(\alpha,\alpha^*)=\gamma\left[-\bar n_{\mathrm{th}}+(\bar n_{\mathrm{th}}+1)N_0+\bar n_{\mathrm{th}}|\alpha|^2\right]-\frac{\gamma(2\bar n_{\mathrm{th}}+1)}{2}(x+y)-i(x-y)\bigl(\omega+\chi(x+y)\bigr)
\end{aligned}
\end{equation}
and the second order term can be written as
\begin{equation}
Q_2(\alpha,\alpha^*)=e^{-(x+y)}\sum_{m,n=0}^{\infty}
\frac{x^m y^n}{m!n!}\left[E_2(m,n)+\frac{1}{2}E_1(m,n)^2\right]
\end{equation}
where
\begin{equation}
E_1(m,n)=-i(\omega-2\chi)k+\frac{\gamma}{2}-\frac{m+n+1}{2}A(k)
+\gamma(\bar n_{\mathrm{th}}+1)N_0+\gamma\bar n_{\mathrm{th}}|\alpha|^2
\end{equation}
\begin{equation}
E_2(m,n)=\frac{m+n+1}{2}\gamma^2\bar n_{\mathrm{th}}(\bar n_{\mathrm{th}}+1)
-\frac{\gamma(\bar n_{\mathrm{th}}+1)}{2}A(k)N_0-\frac{\gamma\bar n_{\mathrm{th}} }{2}A(k)|\alpha|^2 
\end{equation}
The various moments $\langle a^p a^{\dagger q}\rangle$ are computed using the formula
\begin{equation}
\langle a^p a^{\dagger q} \rangle =\frac{1}{\pi} \int d^2\alpha \; \alpha^p (\alpha^*)^q \; Q(\alpha,\alpha^*)
\end{equation}
\subsubsection*{Computation of moments for $T\neq 0$ }
Using the  following identity $\frac{1}{\pi}\int d^2\alpha\,
\alpha^p(\alpha^*)^q e^{-|\alpha-\alpha_0|^2}
=\sum_{r=0}^{\min(p,q)}\binom{p}{r}\binom{q}{r}r!\,
\alpha_0^{p-r}(\alpha_0^*)^{q-r}$ we obtain after some algebra the moments up to (and including) order $t^2$
\begin{align}
\langle a(t)\rangle
&=\alpha_0\left[1+A_1 t+\frac{1}{2}A_2 t^2\right]+O(t^3),\\
\langle a^2(t)\rangle
&=\alpha_0^2\left[1+B_1 t+\frac{1}{2}B_2 t^2\right]+O(t^3),\\
N(t)\equiv\langle a^\dagger(t)a(t)\rangle
&=N_0-\gamma\left(N_0-\bar n_{\mathrm{th}}\right)t
+\frac{\gamma^2}{2}\left(N_0-\bar n_{\mathrm{th}}\right)t^2+O(t^3).
\end{align}
with $A_{i},B_{i}$ given explicitly in Table \ref{tab:moment-coefficients}. 
\begin{table}[t!]
\label{table}
\centering
\small
\renewcommand{\arraystretch}{1.45}
\setlength{\tabcolsep}{8pt}
\begin{tabular}{c@{\qquad}l}
\hline\hline
\textbf{Coeff.} & \textbf{Expression} \\
\hline

$A_1$
&
$\displaystyle
-\frac{\gamma}{2}
-i\left(\omega+2\chi N_0\right)$
\\[3pt]

$A_2$
&
$\displaystyle
-\left(\omega+2\chi N_0\right)^2
-4\chi^2N_0
+\frac{\gamma^2}{4}
+i\gamma\omega
+4i\gamma\chi(N_0-\bar n_{\mathrm{th}})$
\\[3pt]

$B_1$
&
$\displaystyle
-\gamma
-i\left[2(\omega+\chi)+4\chi N_0\right]$
\\[3pt]

$B_2$
&
$\displaystyle
-\left[2(\omega+\chi)+4\chi N_0\right]^2
-16\chi^2N_0
+\gamma^2
+4i\gamma\omega
+i\gamma\chi(12N_0-12\bar n_{\mathrm{th}}+4)$
\\

\hline\hline
\end{tabular}
\caption{Short-time coefficients entering $\langle a(t)\rangle$ and $\langle a^2(t)\rangle$, with $N_0=|\alpha_0|^2$.}
\label{tab:moment-coefficients}
\end{table}
The two combinations entering $V_-(t)$ are
\begin{equation}\label{eq:2pieceVarianceTnonzero}
\langle a^\dagger a\rangle-|\langle a\rangle|^2=
\gamma\bar n_{\mathrm{th}}\,t+\left(4\chi^2N_0^2-\frac{\gamma^2\bar n_{\mathrm{th}}}{2}\right)t^2+O(t^3)
\end{equation}
\begin{equation}
\langle a^2\rangle-\langle a\rangle^2=\alpha_0^2\left[-2i\chi\,t+
\left(-4\chi\omega-2\chi^2(6N_0+1)+2i\gamma\chi(1-\bar n_{\mathrm{th}})\right)t^2\right]
+O(t^3)
\end{equation}
Therefore, 
\begin{equation}\label{eq:1pieceVarianceTnonzero}
\left|\langle a^2\rangle-\langle a\rangle^2\right|=2|\chi|N_0\,t
-2\gamma|\chi|N_0(1-\bar n_{\mathrm{th}})t^2+O(t^3)
\end{equation}
Inserting equations \eqref{eq:2pieceVarianceTnonzero} and \eqref{eq:1pieceVarianceTnonzero} into \eqref{eq:definitionMinimalVariance} we end up with the main result \eqref{vminustemp}.
\newpage

\bibliographystyle{unsrt}
\bibliography{references_local}

\end{document}